\documentclass[preprint,1p,sort&compress]{elsarticle}

\usepackage{hyperref}
\usepackage{graphicx}
\usepackage{amssymb,amsmath,bm}
\usepackage{textcomp}
\usepackage{multirow}
\usepackage{arydshln}
\usepackage{todonotes}
\usepackage{tipa}
\usepackage{rotating}
\usepackage{multicol}
\usepackage[section]{placeins}
\usepackage{tabto}

\journal{Journal of Computer Speech and Language}

\begin{document}

\begin{frontmatter}

\title{Comparing heterogeneous visual gestures for measuring the diversity of visual speech signals}

\author[cvssp]{Helen L Bear\corref{cor1}}
\address[cvssp]{CVSSP, Dept of Electrical Engineering, University of Surrey, Guildford, GU2 7JP, UK}
 \cortext[cor1]{Corresponding author}
\ead{dr.bear@tum.de}

\author[uea]{Richard Harvey}
\address[uea]{School of Computing Sciences, University of East Anglia, Norwich, NR4 7TJ, UK}
\ead{r.w.harvey@uea.ac.uk}

\begin{abstract}
Visual lip gestures observed whilst lipreading have a few working definitions, the most common two are; `the visual equivalent of a phoneme' and `phonemes which are indistinguishable on the lips'. To date there is no formal definition, in part because to date we have not established a two-way relationship or mapping between visemes and phonemes. 
Some evidence suggests that visual speech is highly dependent upon the speaker. So here, we use a phoneme-clustering method to form new phoneme-to-viseme maps for both individual and multiple speakers. We test these phoneme to viseme maps to examine how similarly speakers talk visually and we use signed rank tests to measure the distance between individuals. We conclude that broadly speaking, speakers have the same repertoire of mouth gestures, where they differ is in the use of the gestures.
\end{abstract}


\end{frontmatter}


Computer lipreading is machine speech recognition from the interpretation of lip motion without auditory support \cite{stafylakis2017combining, bear2017bmvcTerms}. There are many motivators for wanting a lipreading machine, for example places where audio is severely hampered by noise such as an airplane cockpit, or where placing a microphone close to a source is impossible such as a busy airport or transport hub \cite{neti2000audio, morade2014lip, bear2014some}

Conventionally, machine lipreading has been implemented on two-dimensional videos filmed in laboratory conditions \cite{7050271,cooke2006audio}. More recently, such datasets have been growing in size to enable deep learning methods to be applied in lipreading systems \cite{chungaccv, thangthai2017interspeech}. Separately there has also been some preliminary work to use depth cameras to capture pose/lip protrusion information \cite{heidenreich2016three, watanabe2016lip} or in the RGB colour space for more discriminative appearance features \cite{rekik2016adaptive}. The challenge with these works are that the results achieved are yet to significantly outperform conventional lipreading systems. The top 1 scores in \cite{chungaccv} are less than \cite{wandicassp2016} and to date the best end-to-end system is that of Stafylakis and Tzimiropoulos who achieved an error rate of $11.29\%$ on a $500$ word vocabulary \cite{stafylakisdeep}.

In developing lipreading systems we know that speech is a bimodal signal, and we use the the visual channel of information for recognition of visual cues or gestures \cite{petridis2013mahnob}. The units within this information channel, in sequence form a signal of its own, but it has no formal definition despite a variety of options presented previously \cite{7074217,hilder2009comparison, chen1998audio, fisher1968confusions, Hazen1027972}. Irrespective of the definition in each paper, these units are commonly referred to as `visemes' and in this paper, we define a viseme as a visual cue (sometimes also referred to as a gesture) that represents a subset of identical phonemes on the lips \cite{bear2015speakerindep, lip_reading18, bear2017bmvcVariability}. This means a set of visemes is always smaller than the set of phonemes \cite{cappelletta2012phoneme}. These visemes are interesting because they help researchers to answer questions about how best to decipher lip motions when affected by issues such as human lipreading \cite{jeffers1971speechreading}, language \cite{newman2012language}, expression \cite{metallinou2010visual}, and camera parameters like resolution \cite{bear2014resolution}. 

Previous work has shown the benefits of deriving speaker-dependent visemes \cite{kricos1982differences,bear2017phoneme} but the cost associated with generating these is significant. Indeed the work by Kricos \cite{kricos1982differences} was limited due to the human subjects required, whereas the data-driven method of Bear \cite{bear2017phoneme} could scale if visual speech ground truths for the test speakers were available in advance. The concept of a unique Phoneme-to-Viseme (P2V) mapping for every speaker is daunting, so here we test the versatility and robustness of speaker-dependent visemes by using the algorithm in \cite{bear2014phoneme} to derive single-speaker, multi-speaker, and multi-speaker-independent visemes and use these in a controlled experiment to answer the following questions; To what extent are such visemes speaker-independent? What is the similarity between these sets of visemes? 

This work is motivated by the many future applications of viseme knowledge. From improving both lipreading and audio-visual speech recognition systems for security and safety, to refereeing sports events and understanding silent films, understanding visual speech gestures has significant future impact on many areas of society. 

In our previous work we investigated isolated word recognition from speaker-dependent visemes \cite{bear2015speakerindep}. Here we extend this to continuous speech. Benchmarked against speaker-dependent results, we experiment with speakers from both the AVLetters2 (AVL2) and Resource Management Audio-Visual (RMAV) datasets. The AVL2 dataset is a dataset of seven utterances per speaker reciting the alphabet. In RMAV the speakers utter continuous speech, sentences from three to six words long for up to 200 sentences each. Our hypothesis is that, with good speaker-specific visemes, we can negate the previous poor performance of speaker independent lipreading. This is because, particularly with continuous speech, information from language and grammar create longer sequences upon which classifiers can discriminate.

The rest of this paper is structured as follows: we discuss the issue of speaker identity in computer lipreading, how this can be a part of the feature extraction method to improve accuracy and how visemes can be generated. We then discuss speaker-independent systems before we introduce the experimental data and methods. We present results on isolated words and continuous speech data. We use the Wilcoxon signed rank \cite{wilcoxon1945individual} to measure the distances between the speaker-dependent P2V maps before drawing conclusions on the observations. 

\section{Speaker-specific visemes}
Speaker appearance, or identity, is known to be important in the recognition of speech from visual-only information (lipreading) \cite{cox2008challenge}, more so than in auditory speech. Indeed appearance data improves lipreading classification over shape only models whether one uses Active Appearance Models (AAM) \cite{bear2014resolution} or Discrete Cosine Tranform (DCT) \cite{heidenreich2016three} features. 

In machine lipreading we have interesting evidence: we can both identify individuals from visual speech information \cite{607030,1703580} and, with deep learning and big data, we have the potential to generalise over many speakers \cite{chungaccv,wand2017improving}. 

One of the difficulties in dealing with visual speech is deciding what the fundamental units for recognition should be. The term {\em viseme} is loosely defined \cite{fisher1968confusions} to mean a visually indistinguishable unit of speech, and a set of visemes is usually defined by grouping together a number of phonemes that have a (supposedly) indistinguishable visual appearance. Several many-to-one mappings from phonemes to visemes have been proposed and investigated \cite{fisher1968confusions}, \cite{lip_reading18}, or \cite{jeffers1971speechreading}. Bear \textit{et al.} showed in \cite{bear2017phoneme} that the best speaker-independent P2V map was devised by Lee \cite{lee2002audio} when recognising isolated words, but for continuous speech a combination of Disney's vowels \cite{disney} and Woodward's \cite{woodward1960phoneme} consonants were better. From this we inferred that language has a significant effect on the appearance of visemes. 

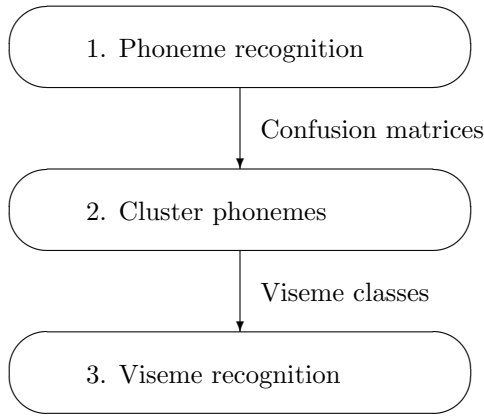
\begin{figure}[!ht] 
\centering 
\setlength{\unitlength}{\columnwidth/100}
\begin{picture}(100, 40)
\put(50,37){\oval(45,8)}
\put(35,36){1. Phoneme recognition}
\put(50,33){\vector(0,-1){8}}
\put(52,28){Confusion matrices}
\put(50,21){\oval(45,8)}
\put(35,20){2. Cluster phonemes}
\put(50,17){\vector(0,-1){8}}
\put(52,12){Viseme classes}
\put(50,5){\oval(45,8)}
\put(35,4){3. Viseme recognition}
\end{picture}
\caption{ Three step process for recognition from visemes. This figure summarizes the process undertaken by Bear et al. in \cite{bear2014phoneme}}
\label{fig:process} 
\end{figure} 

The question then arises to what extent such maps are independent of the speaker, and if so, how speaker independence might be examined. In particular, we are interested in the interaction between the data used to train the models and the viseme classes themselves. 
 
More than in auditory speech, in machine lipreading, speaker identity is important for accurate classification \cite{cox2008challenge}. We know a major difficulty in visual speech is the labeling of classifier units so we need to address the questions; to what extent are such maps independent of the speaker? And if so, how might speaker dependent sets of visemes be examined? Alongside of this, it would be useful to understand the interactions between the model training data and the classes. Therefore in this section we will use both the AVL2 dataset \cite{cox2008challenge} and the RMAV dataset \cite{lan2010improving} to train and test classifiers based upon a series of P2V mappings. 
 
\subsection{Speaker-independence} 
Currently, robust and accurate machine lipreading performances are achieved with speaker-dependent classification models \cite{chungaccv}, this means the test speaker must be included within the classifier training data. A classification model which is trained without the test speaker performs poorly \cite{rashidinvestigation, cox2008challenge}. Thus speaker independence is the ability to classify a speaker who is not involved in the classifier training \cite{bear2017bmvcTerms}. This is a difficult, and as yet, unsolved problem. 

One could wonder if, with a large enough dataset with a significant number of speakers, then it could be sufficient to train classifiers which are generalised to cover a whole population including independent speakers. But we still struggle without a dataset of the size needed to test this theory, particularly as we do not know how much is `enough' data or speakers. Works such as \cite{wand} use domain adaptation \cite{ganin2015unsupervised}, and \cite{improveVis} use Feature-space Maximum Likelihood Linear Regression (fMLLR) features \cite{miao2014improvements,rahmani2017lip}. These achieve significant improvements on previous speaker independent results but still do not match those of speaker dependent accuracy. 

An example of a study into speaker independence in machine lipreading is \cite{cox2008challenge}, here the authors also use the AVL2 dataset and they compare single speaker, multi-speaker and speaker independent classification using two types of classifiers (Hidden Markov Models (HMM) \& Sieves, sieves are a kind of visual filter \cite{bangham1996nonlinear}). However, this investigation uses word labels for classifiers and we are interested to know if the results could be improved using speaker-dependent visemes. 

\section{Description of datasets}
\begin{figure}[!ht]
\centering
\includegraphics[width=0.85\textwidth]{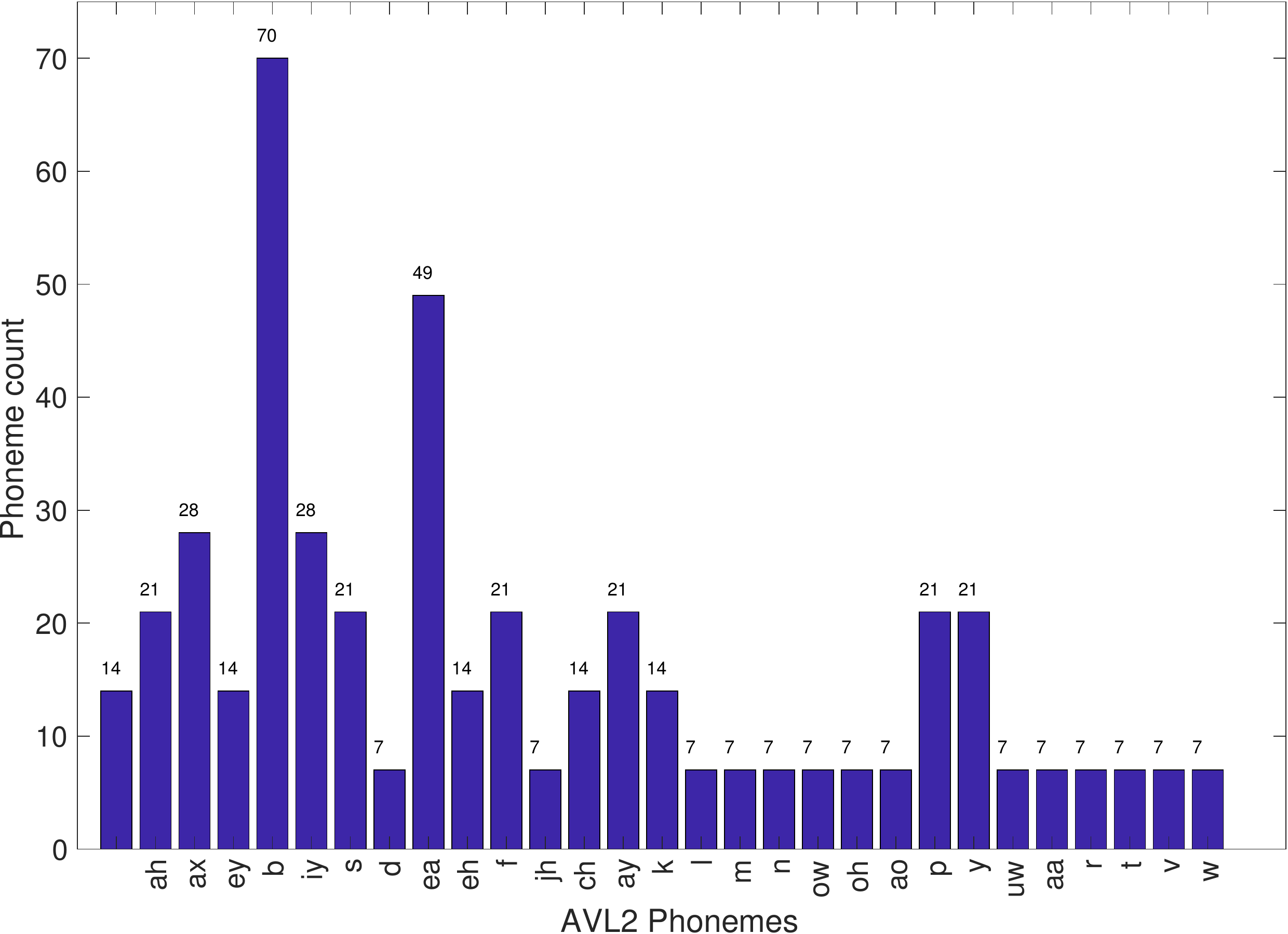}
\caption{Phoneme histogram of AVLetters-2 dataset}
\label{fig:histogram}
\end{figure}

 We use the AVL2 dataset \cite{cox2008challenge}, to train and test recognisers based upon the speaker-dependent mappings. This dataset consists of four British-English speakers reciting the alphabet seven times. The full-faces of the speakers are tracked using Linear Predictors \cite{ong2011robust} and Active Appearance Models \cite{Matthews_Baker_2004} are used to extract lip-only combined shape and appearance features. We select AAM features because they are known to out-perform other feature methods in machine visual-only lipreading \cite{cappelletta2012phoneme}.
Figure~\ref{fig:histogram} shows the count of the 29 phonemes that appear in the phoneme transcription of AVL2, allowing for duplicate pronunciations, (with the silence phoneme omitted). The British English BEEP pronunciation dictionary \cite{beep} is used throughout these experiments.

\begin{figure}[!ht] 
\centering 
\includegraphics[width=.85\textwidth]{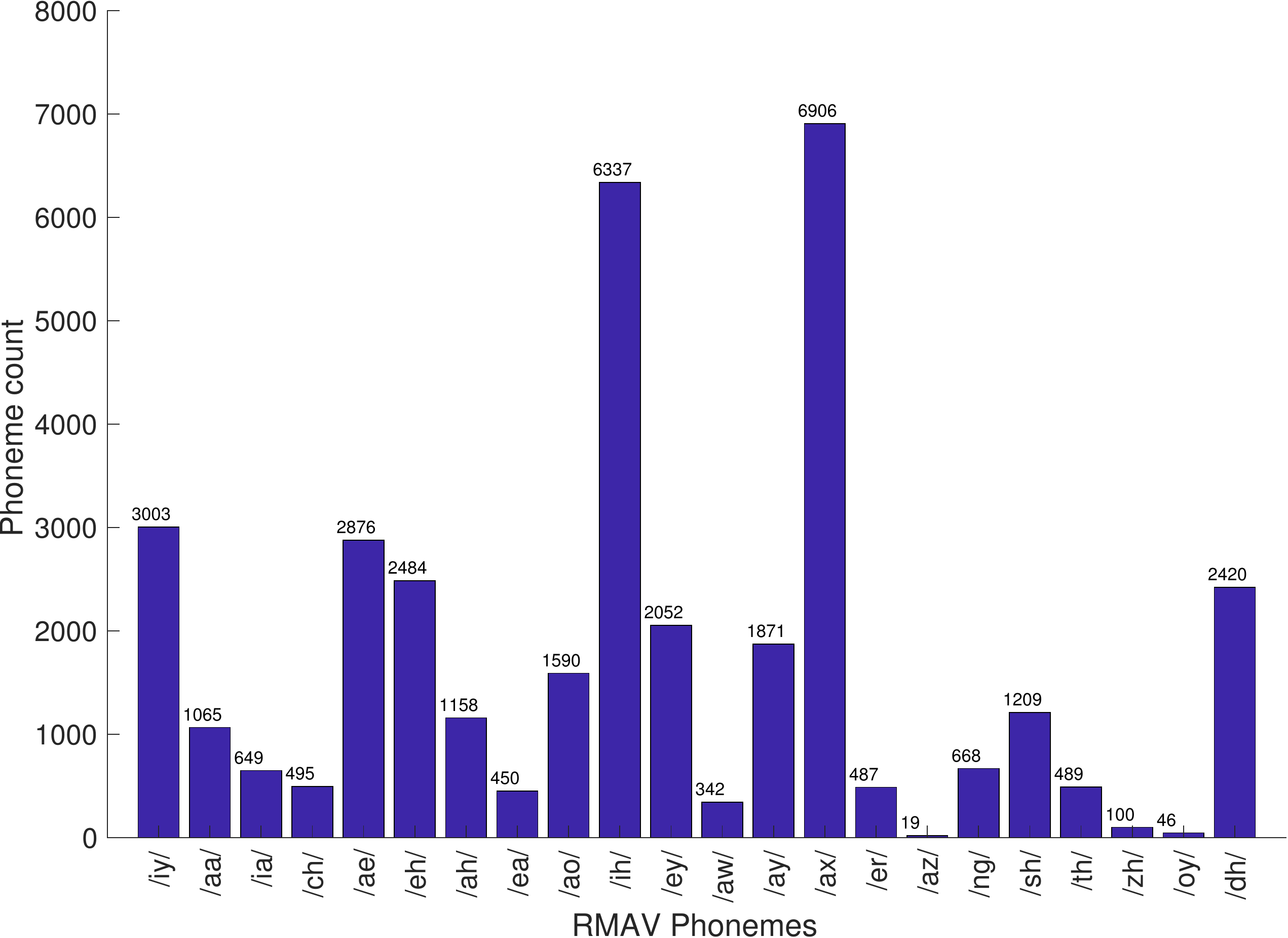} \\
\includegraphics[width=.85\textwidth]{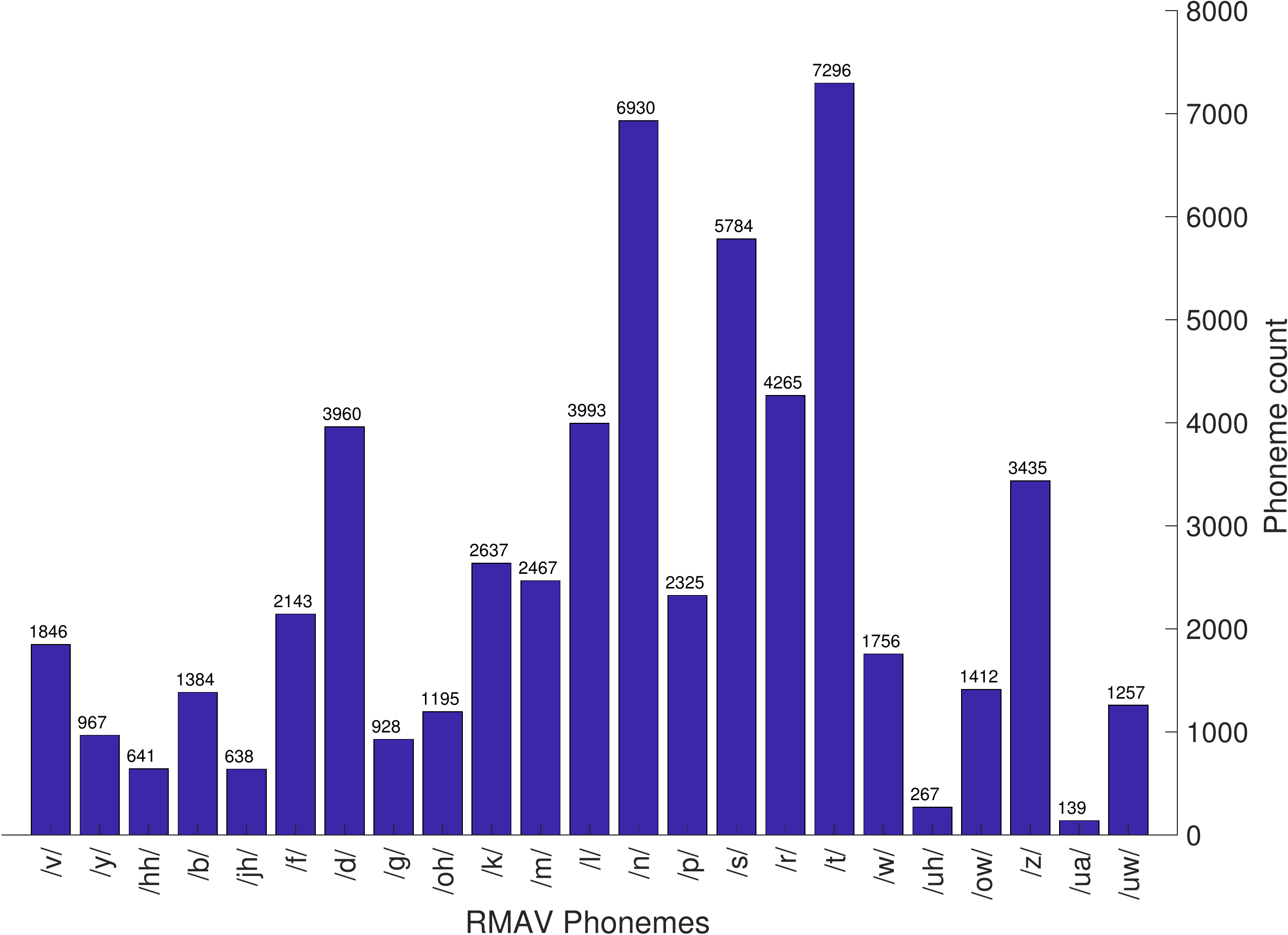}
\caption{Occurrence frequency of phonemes in the RMAV dataset.} 
\label{fig:rmavhistogram} 
\end{figure}

Our second data set is continuous speech. Formerly known as LiLIR, the RMAV dataset consists of $20$ British English speakers (we use $12$, seven male and five female), up to $200$ utterances per speaker of the Resource Management (RM) sentences from \cite{fisher1986darpa} which totals around $1000$ words each. It should be noted the sentences selected for the RMAV speakers are a significantly cut down version of the full RM dataset transcripts. They were selected to maintain as much coverage of all phonemes as possible as shown in Figure~\ref{fig:rmavhistogram} \cite{improveVis}. The original videos were recorded in high definition ($1920 \times 1080$) and in a full-frontal position. 

\FloatBarrier
\subsection{Linear predictor tracking}
Linear predictors have been successfully used to track objects in motion, for example \cite{matas2006learning}. Here linear Predictors are a person-specific and data-driven facial tracking method \cite{sheerman2013non} used for observing visual changes in the face during speech, linear predictor tracking methods have shown robustness that make it possible to cope with facial feature configurations not present in the training data \cite{ong2011robust} by treating each feature independently. 
 
A linear predictor is a single point on or near the lips around which support pixels are used to identify the change in position of the central point between video frames. The central points are a set of single landmarks on the outline of speaker lips. In this method both the lip shape (comprised of landmarks) and the pixel information surrounding the linear predictor positions are intrinsically linked, \cite{ong2008robust}. 

\subsection{Active appearance model features}
\begin{figure}[!h] 
\centering 
\begin{tabular}{l r}
\includegraphics[width=0.45\textwidth]{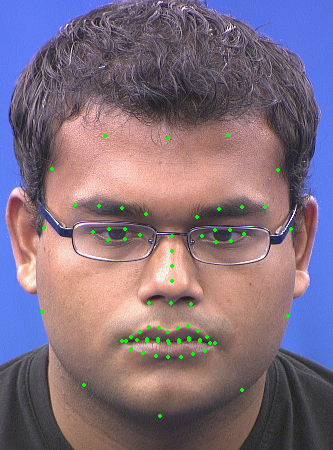} &
\includegraphics[width=0.45\textwidth]{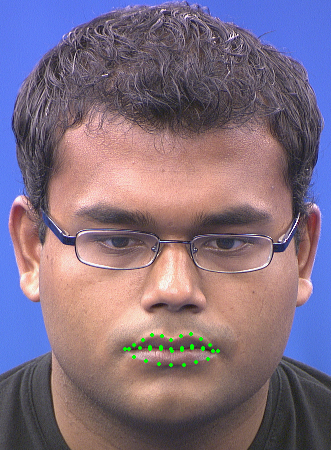}
\end{tabular}
\caption{Example Active Appearance Model shape mesh (left), a lips only model is on the right. Landmarks are in green.} 
\label{fig:egshape} 
\end{figure}

Individual speaker AAM features \cite{Matthews_Baker_2004} of concatenated shape and appearance information have been extracted. The shape features (\ref{eq:shapecombined}) are based solely upon the lip shape and positioning during the speaker speaking e.g. the landmarks in Figure~\ref{fig:egshape} (right) where there are $76$ landmarks in the full face (left) and $34$ landmarks which are modeling the inner and outer lip contours.

The landmark positions can be compactly represented using a linear model of the form:

\begin{equation}
s = s_0 + \sum_{i=1}^ms_ip_i
\label{eq:shapecombined}
\end{equation}
where $s_0$ is the mean shape and $s_i$ are the modes. The appearance features are computed over pixels, the original images having been warped to the mean shape. So $A_0(x)$ is the mean appearance and appearance is described as a sum over modal appearances:
\begin{equation} 
A(x) = A_0(x) + \sum_{i=1}^l{\lambda}_iA_i(x) \qquad \forall x \in S_0
\label{eq:appcombined}
\end{equation} 

Combined features are the concatenation of Shape and Appearance, the AAM parameters of the four AVL2 speakers the twelve RMAV speakers are in Table~\ref{tab:feature_parameters}.

 \begin{table}[!h] 
\centering
\caption{Number of parameters in shape, appearance and combined shape \& appearance AAM features per speaker in AVL2 and RMAV} 
\begin{tabular}{| l | r | r | r |} 
\hline 
Speaker	& Shape & Appearance & Concatenated \\ 
\hline \hline
\multicolumn{4}{|c|}{AVL2} \\
S1	& 11 & 27	& 38 \\ 
S2 	& 9 & 19 & 28 \\ 
S3 	& 9 & 17 & 25 \\ 
S4	& 9 & 17 & 25 \\ 
\hdashline
\multicolumn{4}{|c|}{RMAV} \\
S1	& 13 & 46	& 59 \\ 
S2 	& 13 & 47 & 60 \\ 
S3	& 13 & 43	& 56 \\ 
S4	& 13 & 47	& 60 \\ 
S5 	& 13 & 45 & 58 \\ 
S6	& 13 & 47	& 60 \\ 
S7 	& 13 & 37 & 50 \\ 
S8 	& 13 & 46 & 59 \\ 
S9	& 13 & 45	& 58 \\ 
S10 	& 13 & 45 & 58 \\ 
S11 	& 14 & 72 & 86 \\ 
S12	& 13 & 45	& 58 \\ 
\hline
\end{tabular} 
\label{tab:feature_parameters} 
\end{table} 

\FloatBarrier
\section{Method overview} 
\label{sec:method}
We used the Bear phoneme clustering approach \cite{bear2014phoneme} to produce a series of speaker-dependent P2V maps. 

In summary the clustering method is as follows: 
\begin{enumerate}
\item Perform speaker-dependent phoneme recognition with recognisers that use phoneme labeled classifiers. 
\item By aligning the phoneme output of the recogniser with the transcription of the word uttered, a confusion matrix for each speaker is produced detailing which phonemes are confused with which others. 
\item Any phonemes which are only correctly recognised as themselves (true positive results) are permitted to be single-phoneme visemes.
\item The remaining phonemes are clustered into groups (visemes) based on the confusions identified in Step 2. Confusion is counted as the sum of both false positives ($FP=N\{p|\hat{p}\}$) and false negatives ($FN=N\{\hat{p}|p\}$), $\forall p \in P$. The clustering algorithm permits phonemes to be grouped into a single viseme, $V$ only if each phoneme has been confused with all the others within $V$. 
\item Consonant and vowel phonemes are not permitted to be mixed within a viseme class. Phonemes can only be grouped once. The result of this process is a P2V map $M$ for each speaker. For further details, see \cite{bear2017phoneme}. 
\item These new speaker-dependent viseme sets are then used as units for visual speech recognition for a speaker.
\end{enumerate}

We present an example to illustrate the results of the phoneme clustering method in Table~\ref{tab:firstpass} for the example confusion matrix in Figure~\ref{fig:demoCMforclustering} \cite{bear2017phoneme}. $/v01/$ is a single-phoneme viseme as it only has true positive results. $/v02/$ is a group of $/p1/$, $/p3/$, and $/p7/$ as these all have confusions with each other. Likewise for $/v03/$ which groups $/p2/$ and $/p4/$. Although $/p5/$ was confused with $/p4/$ it was not mixed with $/p2/$ at all so it remains a viseme class of its own, $/v04/$. 

\begin{table}[!h] 
\caption{Example confusion matrix showing confusions between phoneme-labeled classifiers to be used for clustering to create new speaker-dependent visemes from \cite{bear2014phoneme}. True positive classifications are shown in red, confusions of either false positives and false negatives are shown in blue. The estimated classes are listed horizontally and the real classes are vertical.} 
\centering 
\begin{tabular}{|l||r|r|r|r|r|r|r|} 
\hline 
& $/p1/$ & $/p2/$ & $/p3/$ & $/p4/$ & $/p5/$ & $/p6/$ & $/p7/$ \\ 
\hline \hline 
$/p1/$ & {\textcolor{red}1} & 0 & 0 & 0 & 0 & 0 & {\textcolor{blue}4} \\ 
$/p2/$ & 0 & {\textcolor{red}0} & 0 & {\textcolor{blue}2} & 0 & 0 & 0 \\ 
$/p3/$ & {\textcolor{blue}1} & 0 & {\textcolor{red}0} & 0 & 0 & 0 & {\textcolor{blue}1} \\ 
$/p4/$ & 0 & {\textcolor{blue}2} & {\textcolor{blue}1} & {\textcolor{red}0} & {\textcolor{blue}2} & 0 & 0 \\ 
$/p5/$ & {\textcolor{blue}3} & 0 & {\textcolor{blue}1} & {\textcolor{blue}1} & {\textcolor{red}1} & 0 & 0 \\ 
$/p6/$ & 0 & 0 & 0 & 0 & 0 & {\textcolor{red}4} & 0 \\ 
$/p7/$ & {\textcolor{blue}1} & 0 & {\textcolor{blue}3} & 0 & 0 & 0 & {\textcolor{red}1} \\ 
\hline 
\end{tabular}%
\label{fig:demoCMforclustering} 
\end{table} 
 
\begin{table}[!h] 
\centering 
\caption{Example cluster P2V map} 
\begin{tabular} {|l|l|} 
\hline 
Viseme & Phonemes \\ 
\hline \hline 
$/v01/$ & $\{/p6/\} $ \\ 
$/v02/$ & $\{/p1/, /p3/, /p7/\}$ \\ 
$/v03/$ & $\{/p2/, /p4/\}$ \\ 
$/v04/$ & $\{/p5/\} $ \\ 
\hline 
\end{tabular} 
\label{tab:firstpass} 
\end{table} 

Our sets of P2V maps are made up of the following: 
\begin{enumerate} 
\item one multi-speaker P2V map using {\em all} speakers' phoneme confusions (per dataset); \\
and for each speaker;
\begin{enumerate}[\qquad]
\item 2. a speaker-dependent P2V map; 
\qquad \item 3. a speaker-independent P2V map using confusions of all {\em other} speakers in the data. 
\end{enumerate} 
\end{enumerate} 

So we made nine P2V maps for AVL2 (four speaker maps for map types one and three, and one multi-speaker map) and $25$ for RMAV ($12$ speaker maps for map types one and three, and one multi-speaker map). P2V maps were constructed using separate training and test data over cross-validation, seven folds for AVL2 and ten folds for RMAV \cite{efron1983leisurely}. 

With the HTK toolkit \cite{htk34} we built HMM classifiers with the viseme classes in each P2V map. HMMs were flat-started with \texttt{HCompV} and re-estimated $11$ times over (\texttt{HERest}). We classified using \texttt{HVite} and with the output of this we ran \texttt{HResults} to obtain scores. The HMMs each had three states each with an associated five-component Gaussian mixture to keep the results comparable to previous work \cite{982900}. 

To measure the performance of AVL2 speakers we noted that a classification network restricts the output to be one of the 26 letters of the alphabet (with the AVL2 dataset). Therefore, a simplified measure of accuracy in this case;
\begin{equation}
 \frac{\mbox{\# words correct}}{\mbox{\# words classified}}
 \label{eq:avl2acc} 
 \end{equation}

For RMAV a bigram word lattice was built with \texttt{HBuild} and \texttt{HLStats}, and performance is scored as Correctness (\ref{eq:corr}), 

\begin{equation}
\centering
C=\frac{N-D-S}{N} \qquad 
\label{eq:corr}
\end{equation}

where $N$ is the total number of labels in the ground truth, $D$ is the number of deletion errors, and $S$ represents the number of substitution errors. 

\FloatBarrier
\section{Experiment design} 
The P2V maps formed in these experiments are designated as: 
\begin{equation} 
M_n(p,q)\quad 
\label{eq2} 
\end{equation} 
This means the P2V map is derived from speaker $n$, but trained using visual speech data from speaker $p$ and tested using visual speech data from speaker $q$. For example, $M_1(2,3)$ would designate the result of testing a P2V map constructed from Speaker 1, using data from Speaker 2 to train the viseme models, and testing on Speaker 3's data. Thus we will create (over both datasets); 16 P2V maps where $n=p=q$, two P2V maps where $n\neq p=q$, and 16 P2V maps where $n\neq p \neq q$. A total of $34$ P2V maps. 

For ease of reading, we provide in Table~\ref{tab:glossary} a glossary of acronyms used to describe our testing methodology.

\begin{table}[!ht]
\centering
\caption{Test method acronyms.}
\begin{tabular}{|l |l|}
\hline
Acronym & Definition \\
\hline \hline
SSD & Single speaker dependent \\
MS & Multi-speaker \\
DSD & Different-speaker dependent \\
DSD\&D & Different-speaker dependent and Data \\
SI & Speaker-independent \\
\hline
\end{tabular}
\label{tab:glossary}
\end{table}

\subsection{Baseline: Same Speaker-Dependent (SSD) maps} 
\label{sec:expSetup} 
For a baseline we select the same speaker-dependent P2V maps as \cite{bear2014phoneme}. The baseline tests are: $M_1(1,1)$, $M_2(2,2)$, $M_3(3,3)$ and $M_4(4,4)$ (the four speakers in AVL2). Tests for RMAV are: $M_1(1,1)$, $M_2(2,2)$, $M_3(3,3)$, $M_4(4,4)$, $M_5(5,5)$, $M_6(6,6)$, $M_7(7,7)$ and $M_8(8,8)$, $M_9(9,9)$, $M_{10}(10,10)$, $M_{11}(11,11)$ and $M_{12}(12,12)$. These tests are Same Speaker-Dependent (SSD) because the same speaker is used to create the map, to train and test the models. Tables~\ref{tab:sd} depicts how these tests are constructed for AVL2 speakers, the premise is identical for the 12 RMAV speakers.

\begin{table}[!ht] 
\centering 
\caption{Same Speaker-Dependent (SSD) experiments for AVL2 speakers. The results from these tests will be used as a baseline.} 
\begin{tabular}{| l | l | l | l |} 
\hline 
\multicolumn{4}{| c |}{Same speaker-dependent (SD)} \\ 
Mapping ($M_n$) & Training data ($p$) & Test speaker ($q$) & $M_n(p,q)$ \\ 
\hline \hline 
Sp1 & Sp1 & Sp1 & $M_1(1,1)$ \\ 
Sp2 & Sp2 & Sp2 & $M_2(2,2)$ \\ 
Sp3 & Sp3 & Sp3 & $M_3(3,3)$ \\ 
Sp4 & Sp4 & Sp4 & $M_4(4,4)$ \\ 
\hline 
\end{tabular} 
\label{tab:sd} 
\end{table} 


All P2V maps are listed in supplementary materials to this paper. We permit a garbage, $/gar/$, viseme which is a cluster of phonemes in the ground truth which did not appear at all in the output from the phoneme classification (step two of section~\ref{sec:method}). Every viseme is listed with its associated mutually-confused phonemes e.g. for AVL2 Speaker 1 SSD, $M_1$, we see $/v01/$ is made up of phonemes \{/\textturnv/, /iy/, /\textschwa\textupsilon/, /uw/\}. We know from the clustering method in \cite{bear2014phoneme} this means in the phoneme classification, all four phonemes \{/\textturnv/, /iy/, /\textschwa\textupsilon/, /uw/\} were confused with the other three in the viseme. We are using the `strictly-confused' method labeled $B2$ from \cite{bear2017phoneme} with split vowel and consonant groupings as these achieved the highest accurate word classification. 

\subsection{Multi-Speaker (MS) maps} 
A multi-speaker (MS) P2V map forms the viseme classifier labels in the first set of experiments. This map is constructed using phoneme confusions produced by {\em all} speakers in each data set. Again, these P2V maps are in the supplementary material. 

For the multi-speaker experiment notation, we substitute in the word `all' in place of a list of all the speakers for ease of reading. Therefore, the AVL2 MS map is tested as follows: $M_{[all]}(1,1)$, $M_{[all]}(2,2)$, $M_{[all]}(3,3)$ and $M_{[all]}(4,4)$: this is explained in Table~\ref{tab:ms} and the RMAV MS map is tested as: $M_{[all]}(1,1)$, $M_{[all]}(2,2)$, $M_{[all]}(3,3)$, $M_{[all]}(4,4)$, $M_{all]}(5,5)$, $M_{[all]}(6,6)$, $M_{[all]}(7,7)$, $M_{[all]}(8,8)$, $M_{[all]}(9,9)$, $M_{[all]}(10,10)$, $M_{[all]}(11,11)$, $M_{[all]}(12,12)$.

\begin{table}[!ht] 
\centering 
\caption{Multi-Speaker (MS) experiments for AVL2 speakers.} 
\begin{tabular}{| l | l | l | l |} 
\hline 
\multicolumn{4}{| c |}{Multi-Speaker (MS)} \\ 
\multicolumn{1}{| c }{Mapping ($M_n$)} & \multicolumn{1}{ c }{Training data ($p$)} & \multicolumn{1}{ c }{Test speaker ($q$)} & \multicolumn{1}{ c |}{$M_n(p,q)$}\\ 
\hline \hline 
Sp[all] & Sp1 & Sp1 & $M_{[all]}(1,1)$ \\ 
Sp[all] & Sp2 & Sp2 & $M_{[all]}(2,2)$ \\ 
Sp[all] & Sp3 & Sp3 & $M_{[all]}(3,3)$ \\ 
Sp[all] & Sp4 & Sp4 & $M_{[all]}(4,4)$ \\ 
\hline 
\end{tabular} 
\label{tab:ms} 
\end{table} 


\subsection{Different Speaker-Dependent maps \& Data (DSD\&D)} 
The second set of tests within this experiment start to look at using P2V maps with different test speakers. This means the HMM classifiers trained on each single speaker are used to recognise data from alternative speakers. 

\begin{table}[!ht] 
\centering 
\caption{Different Speaker-Dependent maps and Data (DSD\&D) experiments with the four AVL2 speakers.} 
\begin{tabular}{| l | l | l | l |} 
\hline 
\multicolumn{4}{| c |}{Different Speaker-Dependent maps \& Data (DSD\&D)} \\ 
Mapping ($M_n$) & Training data ($p$) & Test speaker ($q$) & $M_n(p,q)$ \\ 
\hline \hline 
Sp2 & Sp2 & Sp1 & $M_2(2,1)$ \\ 
Sp3 & Sp3 & Sp1 & $M_3(3,1)$ \\ 
Sp4 & Sp4 & Sp1 & $M_4(4,1)$ \\ 
Sp1 & Sp1 & Sp2 & $M_1(1,2)$ \\ 
Sp3 & Sp3 & Sp2 & $M_3(3,2)$ \\ 
Sp4 & Sp4 & Sp2 & $M_4(4,2)$ \\ 
Sp1 & Sp1 & Sp3 & $M_1(1,3)$ \\ 
Sp2 & Sp2 & Sp3 & $M_2(2,3)$ \\ 
Sp4 & Sp4 & Sp3 & $M_4(4,3)$ \\ 
Sp1 & Sp1 & Sp4 & $M_1(1,4)$ \\ 
Sp2 & Sp2 & Sp4 & $M_2(2,4)$ \\ 
Sp3 & Sp3 & Sp4 & $M_3(3,4)$ \\ 
\hline 
\end{tabular} 
\label{tab:sid} 
\end{table} 

Within AVL2 this is completed for all four speakers using the P2V maps of the other speakers, and the data from the other speakers. Hence for Speaker 1 we construct $M_2(2,1)$, $M_3(3,1)$ and $M_4(4,1)$ and so on for the other speakers, this is depicted in Table~\ref{tab:sid}. 

 
 For the RMAV speakers, we undertake this for all $12$ speakers using the maps of the $11$ others. 
In this set of tests we are replicating the format of \cite{bear2015speakerindep} where $p \neq q$ but we use speaker-dependent visemes to mitigate the effect of speaker independence between training and test data. 
 
\subsection{Different Speaker-Dependent maps (DSD)} 
Now we wish to isolate the effects of the HMM classifier from the effect of using different speaker dependent P2V maps by training the classifiers on single speakers with the labels of the alternative speaker P2V maps. E.g. for AVL2 Speaker $1$, the tests are: $M_2(1,1)$, $M_3(1,1)$ and $M_4(1,1)$. (All tests are listed in Table~\ref{tab:si}). 

\begin{table}[!h] 
\centering 
\caption{Different Speaker-Dependent maps (DSD) experiments for AVL2 speakers.} 
\begin{tabular}{| l | l | l | l |} 
\hline 
\multicolumn{4}{| c |}{Different Speaker-Dependent maps (DSD)} \\ 
Mapping ($M_n$) & Training data ($p$) & Test speaker ($q$) & $M_n(p,q)$ \\ 
\hline \hline 
Sp2 & Sp1 & Sp1& $M_2(1,1)$ \\ 
Sp3 & Sp1 & Sp1 & $M_3(1,1)$ \\ 
Sp4 & Sp1 & Sp1 & $M_4(1,1)$ \\ 
Sp1 & Sp2 & Sp2 & $M_1(2,2)$ \\ 
Sp3 & Sp2 & Sp2 & $M_3(2,2)$ \\ 
Sp4 & Sp2 & Sp2 & $M_4(2,2)$ \\ 
Sp1 & Sp3 & Sp3 & $M_1(3,3)$ \\ 
Sp2 & Sp3 & Sp3 & $M_2(3,3)$ \\ 
Sp4 & Sp3 & Sp3 & $M_4(3,3)$ \\ 
Sp1 & Sp4 & Sp4 & $M_1(4,4)$ \\ 
Sp2 & Sp4 & Sp4 & $M_2(4,4)$ \\ 
Sp3 & Sp4 & Sp4 & $M_3(4,4)$ \\ 
\hline 
\end{tabular} 
\label{tab:si} 
\end{table} 

These are the same P2V maps as in our SSD baseline but trained and tested differently. 


\subsection{Speaker-Independent maps (SI)} 
Finally, the last set of tests looks at speaker independence in P2V maps. Here we use maps which are derived using all speakers confusions bar the test speaker. This time we substitute the symbol `$\neg x$' in place of a list of speaker identifying numbers, meaning `not including speaker $x$'. The tests for these maps are as follows $M_{\neg 1}(1,1)$, $M_{\neg 2}(2,2)$, $M_{\neg 3}(3,3)$ and $M_{\neg 4}(4,4)$ as shown in Table~\ref{tab:sim} 
for AVL2 
speakers. Speaker independent P2V maps for all speakers are in this papers supplementary materials

\begin{table}[!ht] 
\centering 
\caption{Speaker-Independent (SI) experiments with AVL2 speakers.} 
\begin{tabular}{| l | l | l | l |} 
\hline 
\multicolumn{4}{| c |}{Speaker-Independent (SI)} \\ 
Mapping ($M_n$) & Training data ($p$) & Test speaker ($q$) & $M_n(p,q)$\\ 
\hline \hline 
Sp$\neg$1 & Sp1 & Sp1 & $M_{\neg 1}(1,1)$ \\ 
Sp$\neg$2 & Sp2 & Sp2 & $M_{\neg 2}(2,2)$ \\ 
Sp$\neg$3 & Sp3 & Sp3 & $M_{\neg 3}(3,3)$ \\ 
Sp$\neg$4 & Sp4 & Sp4 & $M_{\neg 4}(4,4)$ \\ 
\hline 
\end{tabular} 
\label{tab:sim} 
\end{table} 
 

\section{Measuring the effects of homophenes} 
Bauman \cite{Joumun2008} suggests we make 13-15 motions per second during normal speech but are only able to pick up eight or nine. Bauman defines these motions which are so visually similar for distinct words they can only be differentiated with acoustic help as homophenes. For example, in the AVL2 data the words are the letters of the alphabet, The phonetic translation of the word `B' is `$/b/ /iy/$' and of `D' is `$/d/ /iy/$'. Using $M_2(2,2)$ to translate these into visemes they are identical `$/v08/ /v01/$'.
 
\begin{table}[!ht] 
\centering 
\caption{Count of homophenes per P2V map} 
\begin{tabular}{| l | r || l | r | } 
\hline
\multicolumn{2}{|c||}{SD Maps} & \multicolumn{2}{|c|}{SI Maps} \\
\hline 
Map 	&	Tokens $T$ & Map & Tokens\\ 
\hline \hline 
$M_1$	&	19	& $M_{!1}$	&	17 \\
$M_2$	&	19	& $M_{!2}$	&	18	\\
$M_3$	& 	24	& $M_{!3}$	&	20	\\ 
$M_4$	& 	24	& $M_{!4} $	&	15	\\
\hdashline 
$M_{[all]}$	&	14	& & \\
\hline 
\end{tabular} 
\label{tab:homophones} 
\end{table} 

Permitting variations in pronunciation, the total number of $T$ tokens (each unique word counts as one token) for each map after each word has been translated to speaker-dependent visemes are listed in Tables~\ref{tab:homophones} and~\ref{tab:homophonesrmav}. More homophenes means a greater the chance of substitution errors and a reduced correct classification. 
We calculate the homophene effect, $H$, as measured in~(\ref{eq:homophone}). Where $T$ is the number of tokens (unique words) and $W$ is the number of total words available in a single speaker's ground truth transcriptions. 

\begin{equation}
H=1-\frac{T}{W}
\label{eq:homophone}
\end{equation}

An example of a homophene are the words `talk' and `dog'. If one uses Jeffers visemes, both of these words transcribed into visemes become `$/C/$ $/V1/$ $/H/$' meaning that recognition of this sequence of visemes, will represent what acoustically are two very distinct words. Thus distinguishing between `talk' and `dog' is impossible, without the use side information such as a word lattice. This is the power of the word network \cite{thangthai2017comparing, bear2018boosting}.

\begin{table}[!ht]
\centering
\caption{Homophenes, $H$ in words, phonemes, and visemes for RMAV}
\begin{tabular}{|l|r|r|r|}
\hline
Speaker & Word & Phoneme & SD Visemes \\
\hline \hline
Sp01 & 0.64157 & 0.64343 & 0.70131\\
Sp02 & 0.72142 & 0.72309 & 0.76693\\
Sp03 & 0.67934 & 0.68048 & 0.73950\\
Sp04 & 0.68675 & 0.68916 & 0.74337\\
Sp05 & 0.48018 & 0.48385 & 0.58517\\
Sp06 & 0.69547 & 0.69726 & 0.74791\\
Sp07 & 0.69416 & 0.69607 & 0.74556\\
Sp08 & 0.69503 & 0.69752 & 0.74907\\
Sp09 & 0.68153 & 0.68280 & 0.73439\\
Sp10 & 0.70146 & 0.70328 & 0.75243 \\
Sp11 & 0.70291 & 0.70499 & 0.75623\\
Sp12 & 0.63651& 0.64317 & 0.70699\\
\hline
\end{tabular}
\label{tab:homophonesrmav}
\end{table}
\FloatBarrier
\section{Analysis of speaker independence in P2V maps} 
\begin{figure}[h] 
\centering 
\includegraphics[width=0.9\linewidth]{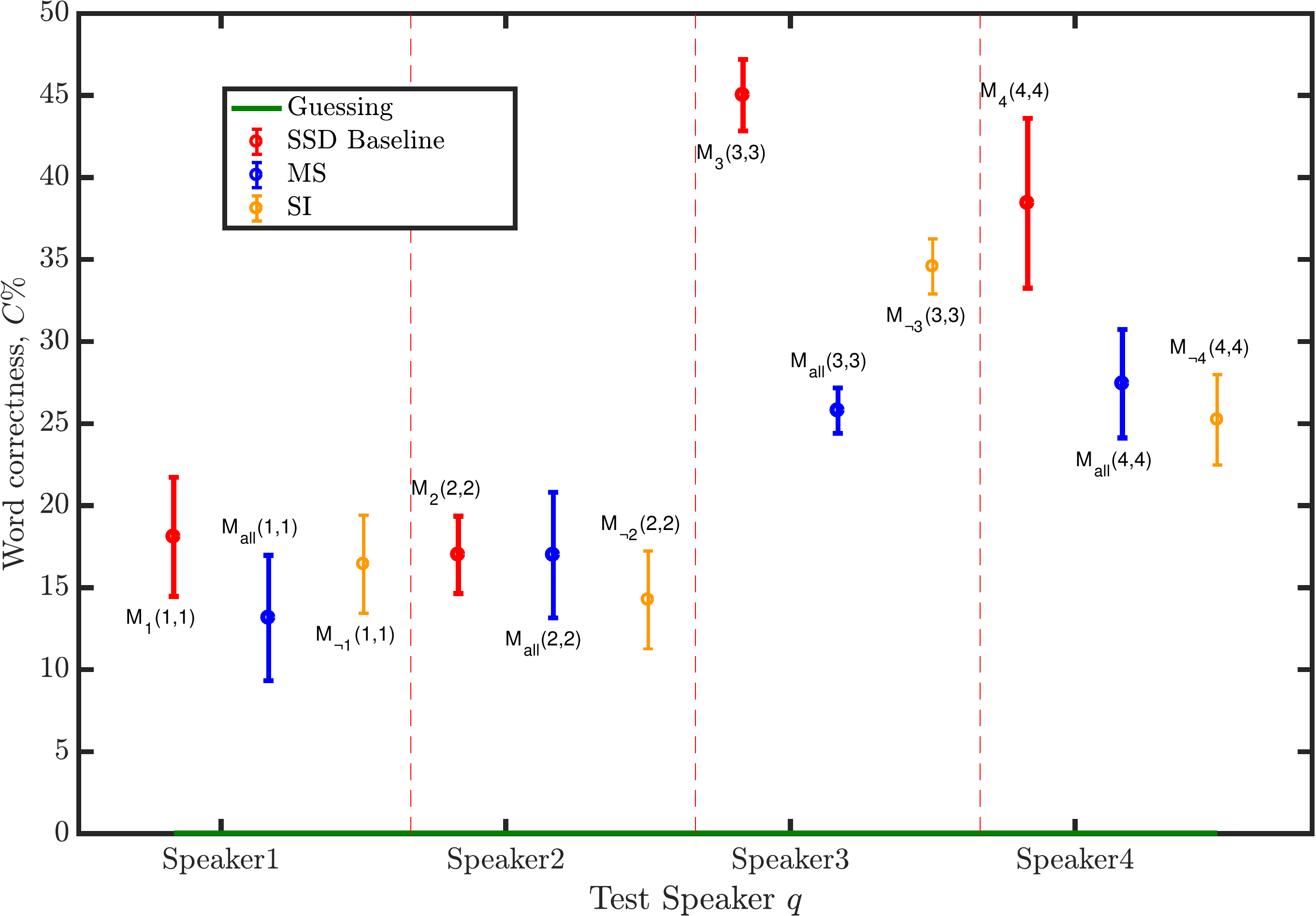} 
\caption{Word correctness, $C\pm1$s.e., using MS and SI P2V maps AVL2} 
\label{fig:accuracy} 
\end{figure}

Figure~\ref{fig:accuracy} shows the correctness of both the MS viseme set (in blue) and the SI tests (in orange) (Tables~\ref{tab:ms} and~\ref{tab:sim}) against the SSD baseline (red) for AVL2 speakers. Word correctness, $C$ is plotted on the $y$-axis. For the MS classifiers, these are all built on the same map $M_{all}$, trained and tested on the same single speaker so, $p=q$. Therefore the tests are: $M_{all}(1,1)$, $M_{all}(2,2)$, $M_{all}(3,3)$, $M_{all}(4,4)$. To test the SI maps, we plot $M_{!1}(1,1)$, $M_{!2}(2,2)$, $M_{!3}(3,3)$ and $M_{!4}(4,4)$. The SSD baseline is on the left of each speakers section of the figure. Note that guessing would give a correctness of $1/N$, where $N$ is the total number of words in the dataset. For AVL2 this is $26$, for RMAV speakers this ranges between $1362$ and $1802$).

There is no significant difference on Speaker 2, and while Speaker 3 word classification is reduced, it is not eradicated. It is interesting for Speaker 3, for whom their speaker-dependent classification was the best of all speakers, the SI map ($M_{!3}$) out performs the multi-speaker viseme classes ($M_{all}$) significantly. This maybe due to Speaker 3 having a unique visual talking style which reduces similaritie.pdfs with Speakers 1, 2 \& 4. But more likely, we see the $/iy/$, phoneme is not classified into a viseme in $M_3$, whereas it is in $M_1$, $M_2$ \& $M_4$ and so re-appears in $M_{all}$. Phoneme $/iy/$ is the most common phoneme in the AVL2 data. This suggests it may be best to avoid high volume of phonemes for deriving visemes as we are exploiting speaker individuality to make better viseme classes.

\begin{figure}[h] 
\centering 
\includegraphics[width=0.9\linewidth]{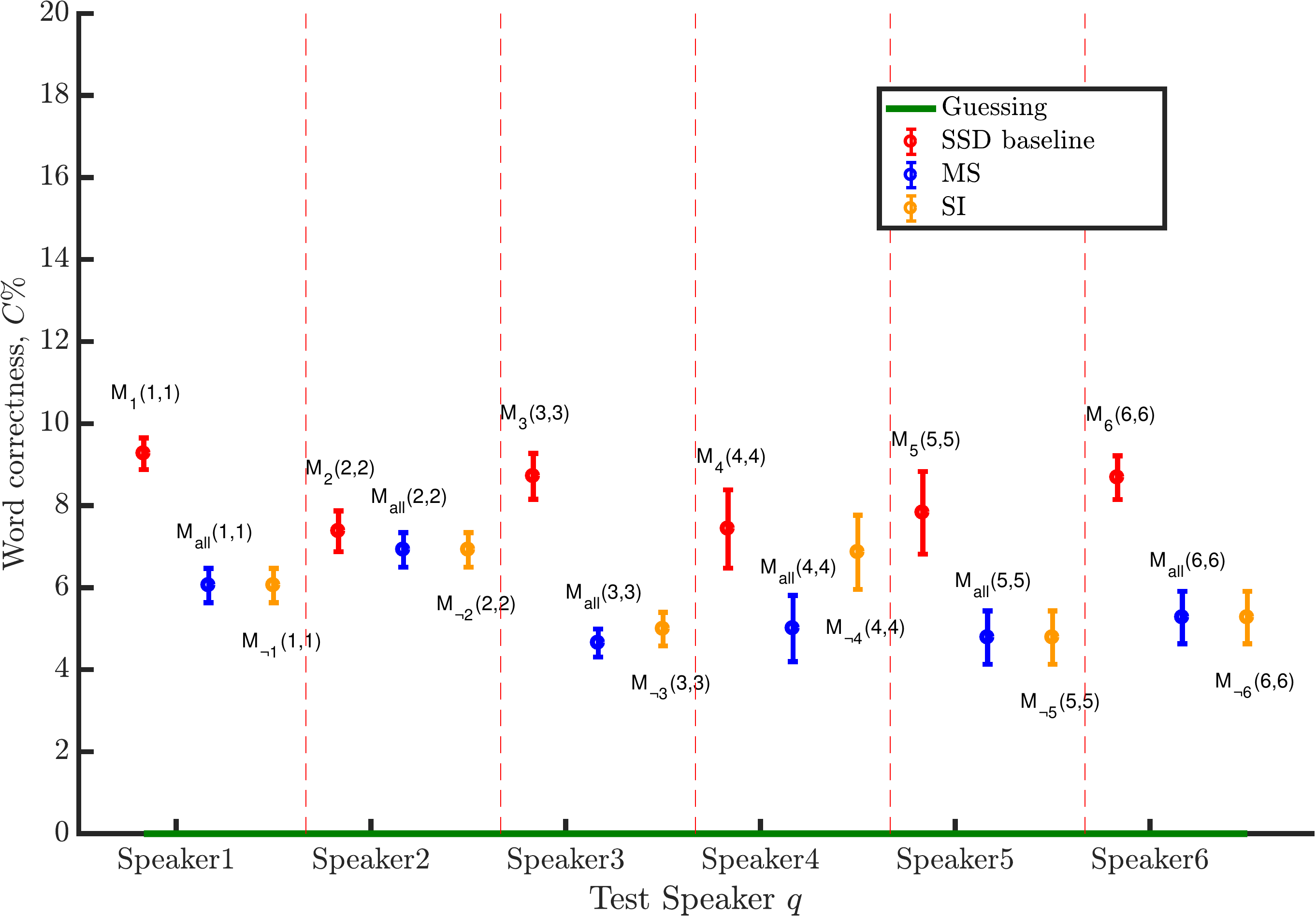} 
\caption{Word correctness, $C\pm1$s.e., using RMAV speakers 1-6 MS and SI P2V maps} 
\label{fig:accuracy1} 
\end{figure}

\begin{figure}[h] 
\centering 
\includegraphics[width=0.9\linewidth]{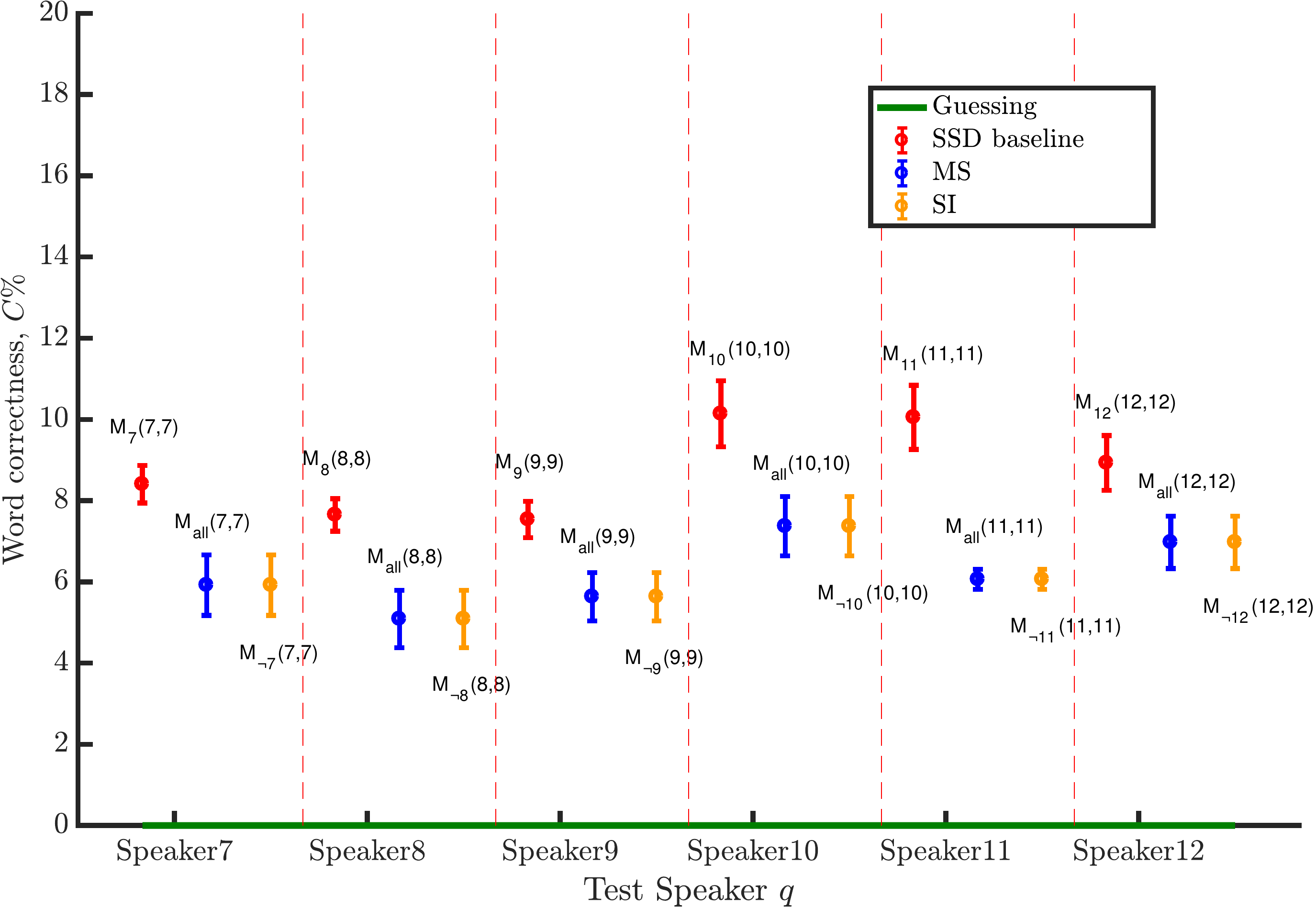} 
\caption{Word correctness, $C\pm1$s.e. using RMAV speakers 7-12 MS and SI P2V maps} 
\label{fig:accuracy2} 
\end{figure}

We have plotted the same MS \& SI experiments on RMAV speakers in Figures~\ref{fig:accuracy1} and~\ref{fig:accuracy2} (six speakers in each figure).
In continuous speech, all but Speaker 2 are significantly negatively affected by using generalized multi-speaker visemes, whether the visemes include the test speakers phoneme confusions or not. This reinforces knowledge of the dependency on speaker identity in machine lipreading but we do see the scale of this effect depends on which two speakers are being compared. For the exception speaker (Speaker 2 in Figure~\ref{fig:accuracy1}) there is only a insignificant decrease in correctness when using MS and SI visemes. Therefore an optimistic view suggests it could be possible with making multi-speaker visemes based upon groupings of visually similar speakers, even better visemes could be created. The challenge remains in knowing which speakers should be grouped together before undertaking P2V map derivation.

\FloatBarrier
\subsection{Different Speaker-Dependent\& Data (DS\&D) results}
\label{sec:dsddResults}
\begin{figure}[h] 
\centering 
\includegraphics[width=0.9\linewidth]{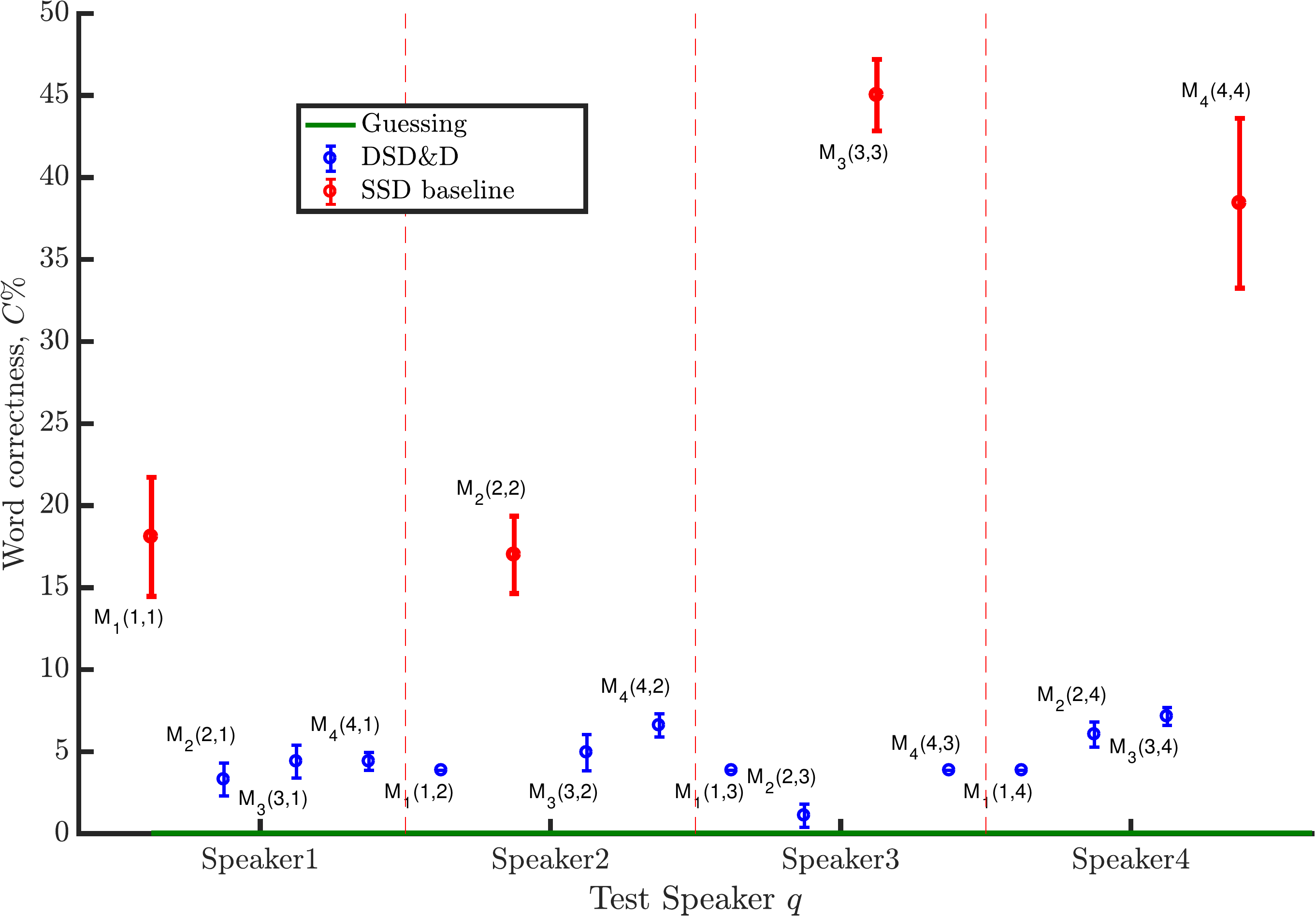} 
\caption{Word correctness, $C\pm1$s.e., of the AVL2 DSD\&D tests Baseline is SSD tests in red.} 
\label{fig:indep_Corr} 
\end{figure}

Figure~\ref{fig:indep_Corr} shows the word correctness of AVL2 speaker-dependent viseme classes on the $y$-axis. Again in this figure, the baseline is $n=p=q$ for all $M$. These are compared to the DSD\&D tests: $M_2(2,1)$, $M_3(3,1)$, $M_4(4,1)$ for Speaker 1, $M_1(1,2)$, $M_3(3,2)$, $M_4(4,2)$ for Speaker 2, $M_1(1,3)$, $M_2(2,3)$, $M_4(4,3)$ for Speaker 3 and $M_1(1,4)$, $M_2(2,4)$, $M_3(3,4)$ for Speaker 4 as in Table~\ref{tab:sid}. 

\begin{figure}[h] 
\centering 
\includegraphics[width=0.9\linewidth]{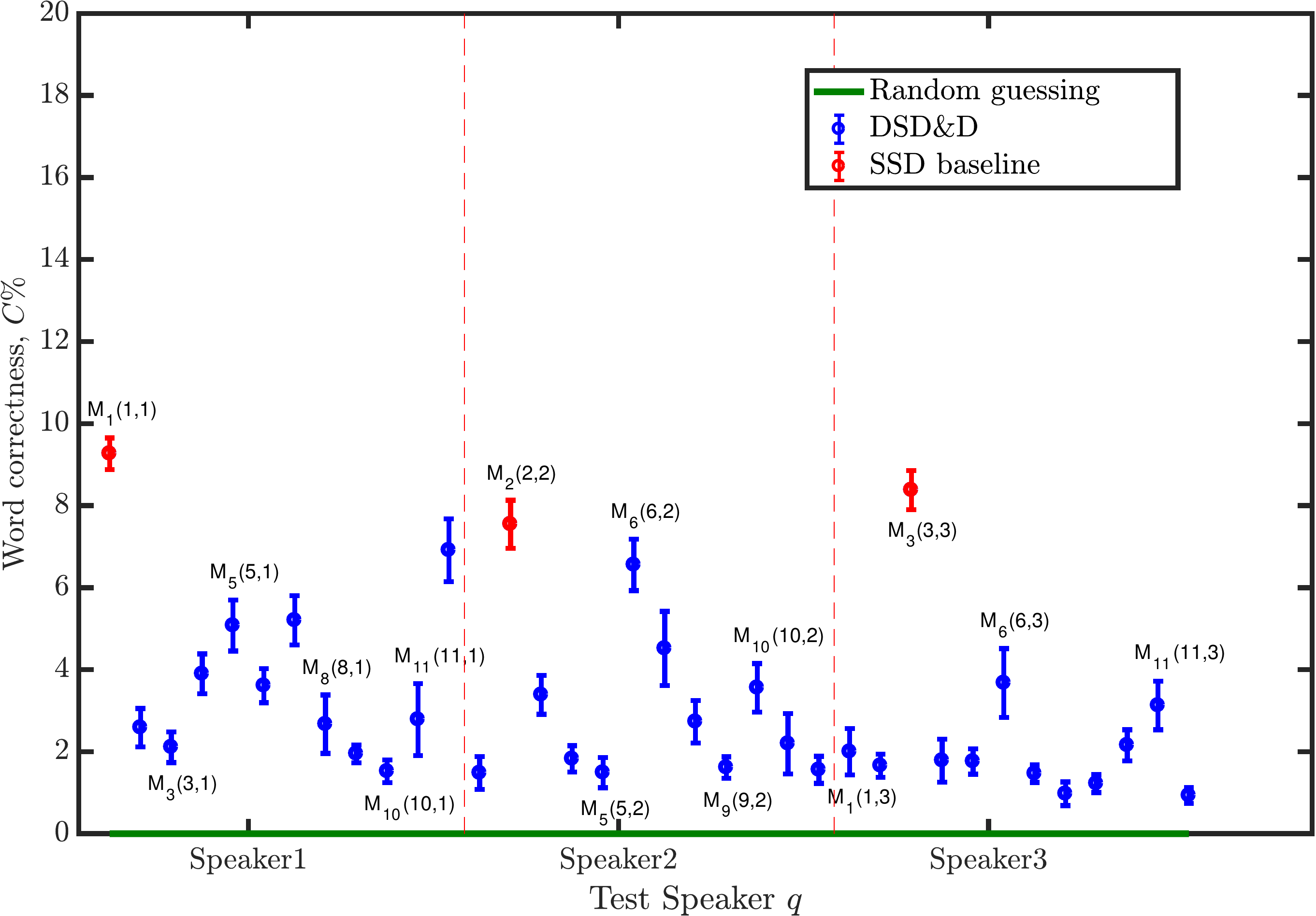} 
\caption{Word correctness, $C\pm1$s.e., of the RMAV speakers 1-3 DSD\&D tests. SSD baseline in red} 
\label{fig:indep_CorrL1} 
\end{figure}

\begin{figure}[h] 
\centering 
\includegraphics[width=0.9\linewidth]{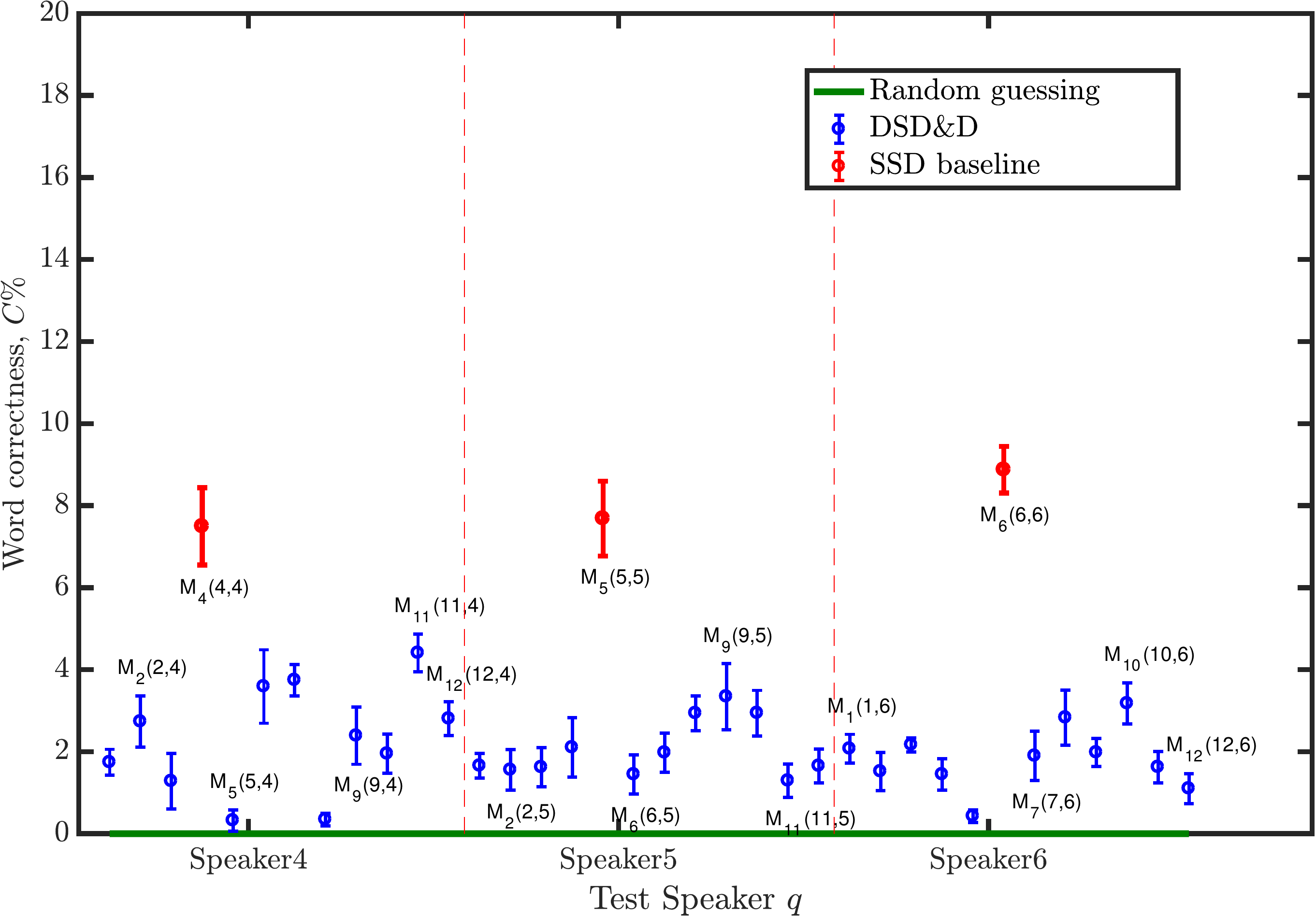} 
\caption{Word correctness, $C\pm1$s.e., of the RMAV speakers 4-6 DSD\&D tests. SSD baseline in red} 
\label{fig:indep_CorrL4} 
\end{figure}

\begin{figure}[h] 
\centering 
\includegraphics[width=0.9\linewidth]{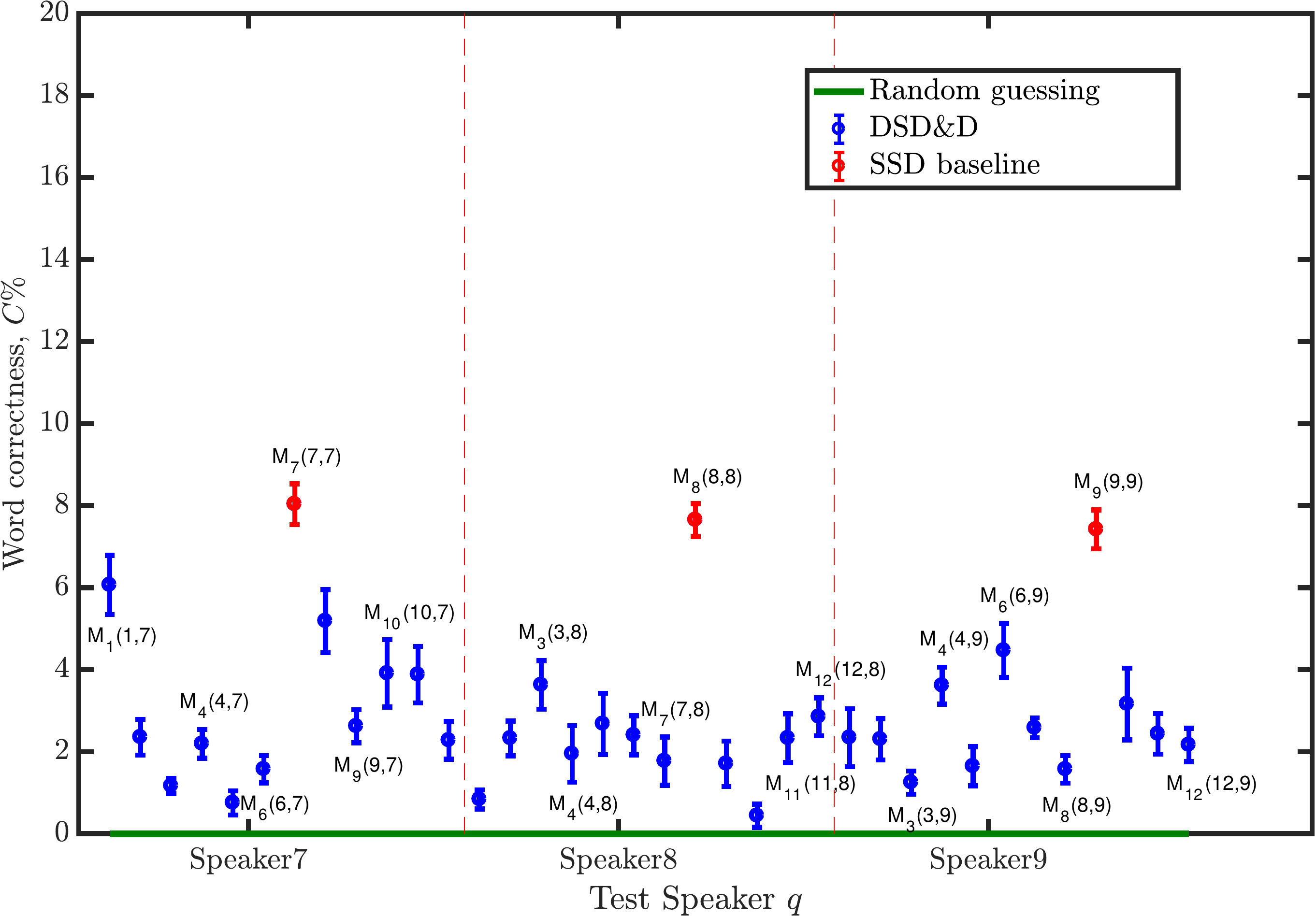} 
\caption{Word correctness, $C\pm1$s.e., of the RMAV speakers 7-9 DSD\&D tests. SSD baseline in red} 
\label{fig:indep_CorrL7} 
\end{figure}

\begin{figure}[h] 
\centering 
\includegraphics[width=0.9\linewidth]{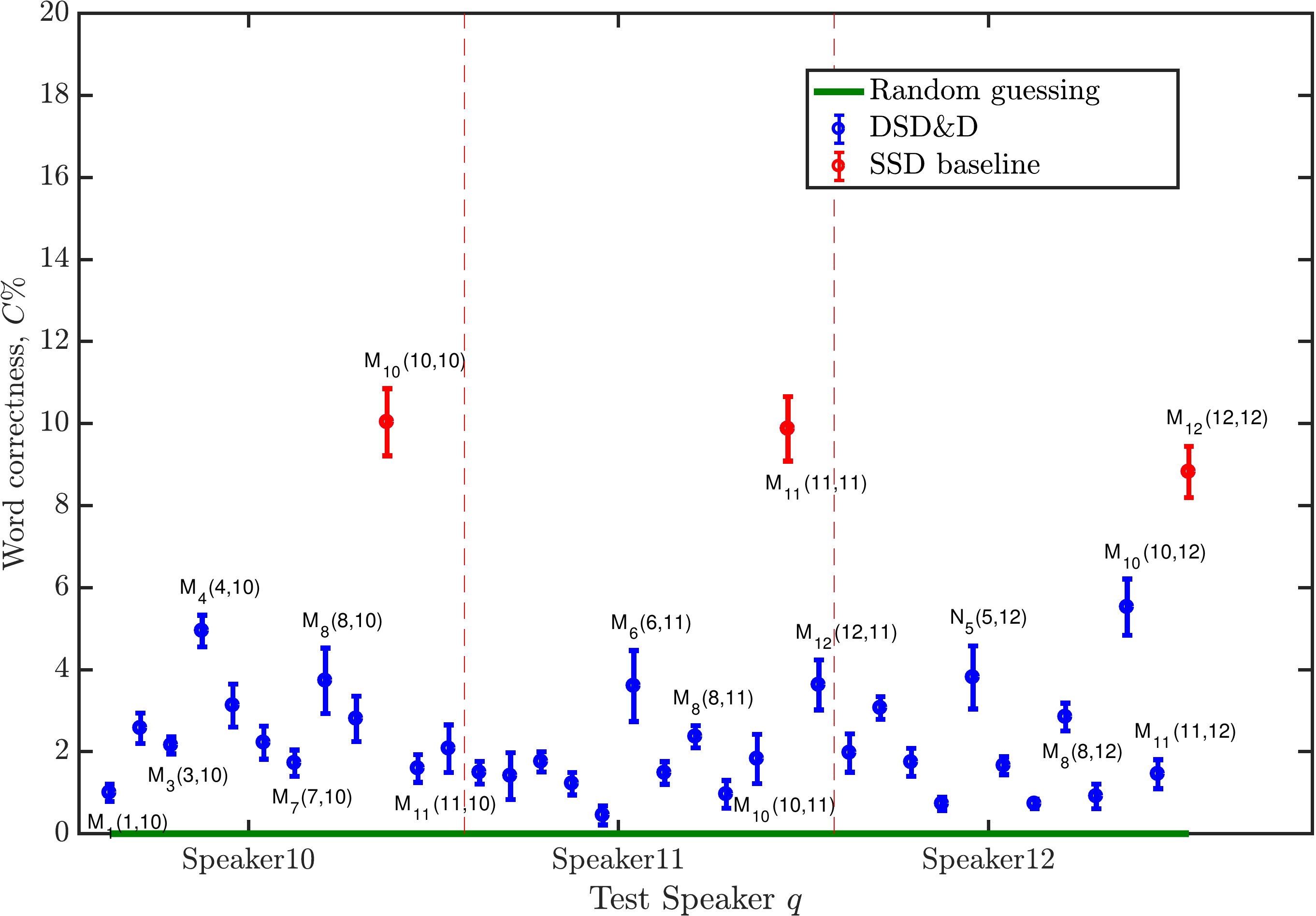} 
\caption{Word correctness, $C\pm1$s.e., of the RMAV speakers 10-12 DSD\&D tests. SSD baseline in red} 
\label{fig:indep_CorrL10} 
\end{figure}

For isolated word classification, DSD\&D HMM classifiers are significantly worse than SSD HMMs, as all results where $p$ is not the same speaker as $q$ are around the equivalent performance of guessing. This correlates with similar tests of independent HMMs in \cite{cox2008challenge}. This gap is attributed to two possible effects, either -- the visual units are incorrect, or they are trained on the incorrect speaker. Figures~\ref{fig:indep_CorrL1}, ~\ref{fig:indep_CorrL4}, ~\ref{fig:indep_CorrL7}, \&~\ref{fig:indep_CorrL10} show the same tests but on the continuous speech data, we have plotted three test speakers per figure. 
 
 
As expected some speakers significantly deteriorate the classification rates when the speaker used to train the classifier is not the same as the test speaker ($p\neq q$). As an example we look at Speaker 1 on the leftmost side of Figure~\ref{fig:indep_CorrL1} where we have plotted Word Correctness for the DSD\&D tests. Here the test speaker is Speaker 1. The speaker-dependent maps for all $12$ speakers have been used to build HMMs classifiers and tested on speaker 1. All $C_w$ for P2V maps significantly reduces except that trained on speaker one. However, in comparison to the AVL2 results, -- this reduction in $C_w$ is not as low as guessing. By capturing language in speaker dependent sets of visemes, we are now less dependent on the speaker identity in the training data. This suggestion is supported by the knowledge of how much of conventional lip reading systems accuracies came from the language model. 

Looking at Figures~\ref{fig:indep_CorrL4} to~\ref{fig:indep_CorrL10} these patterns are consistent. The exception is speaker $2$ in Figure~\ref{fig:indep_CorrL1} where we see that by using the map of speaker 10, $M_{10}$ we do not experience a significant decrease in $C_w$. Furthermore, if we look at Speaker 10's results in Figure~\ref{fig:indep_CorrL10}, all other P2V maps negatively affect speaker 10's $C_w$. This suggest that adaptation between speakers may be directional, that is, we could lipread Speaker 2 having trained on Speaker 10, but not vice versa.

\subsubsection{Continuous speech gestures, or isolated word gestures?}
If we compare these figures to the isolated words results \cite{bear2015speakerindep}, either the extra data in this larger data set or the longer sentences in continuous speech have made a difference. 
Table~\ref{tab:diffs} lists the differences for all speakers on both datasets and the difference between isolated words and continuous speech is between $3.83\%$ to $37.74\%$. Furthermore, with isolated words, the performance attained by speaker-independent tests was shown in cases to be worse than guessing. Whilst the poorest P2V maps might be low, they are all significantly better than guessing regardless of the test speakers.

\begin{table}[!ht]
\centering
\caption{Correctness $C$ with AVL2 and RMAV speakers} 
\begin{tabular}{|l |r |r |r |r |}
\hline \noalign{\smallskip}
\multirow{2}{*}{AVL2}	 & Sp1 	& Sp2 	& Sp3 	& Sp4 	 \\
 	& 14.06 	& 11.87	& 42.08	& 32.75 \\
\hline
\end{tabular}

\resizebox{\columnwidth}{!}{%
\begin{tabular}{|l |r |r |r |r |r |r |r |r |r |r |r |r |}
\hline \noalign{\smallskip}
\multirow{2}{*}{RMAV}	 & Sp1 	& Sp2 	& Sp3 	& Sp4 	& Sp5 	& Sp6 	& Sp7 	& Sp8 	& Sp9 	& Sp10 	& Sp11 	& Sp12 \\

 	& 5.78	& 4.74	& 6.49 	& 5.13 	& 5.57 	& 4.92 	& 6.60 	& 5.19	& 5.64 	& 7.03	& 7.49 	& 8.04 \\
\hline
\end{tabular} %
}
\label{tab:diffs}
\end{table}

\FloatBarrier
\subsection{DSD Results}
\begin{figure}[!ht] 
\centering 
\includegraphics[width=0.9\linewidth]{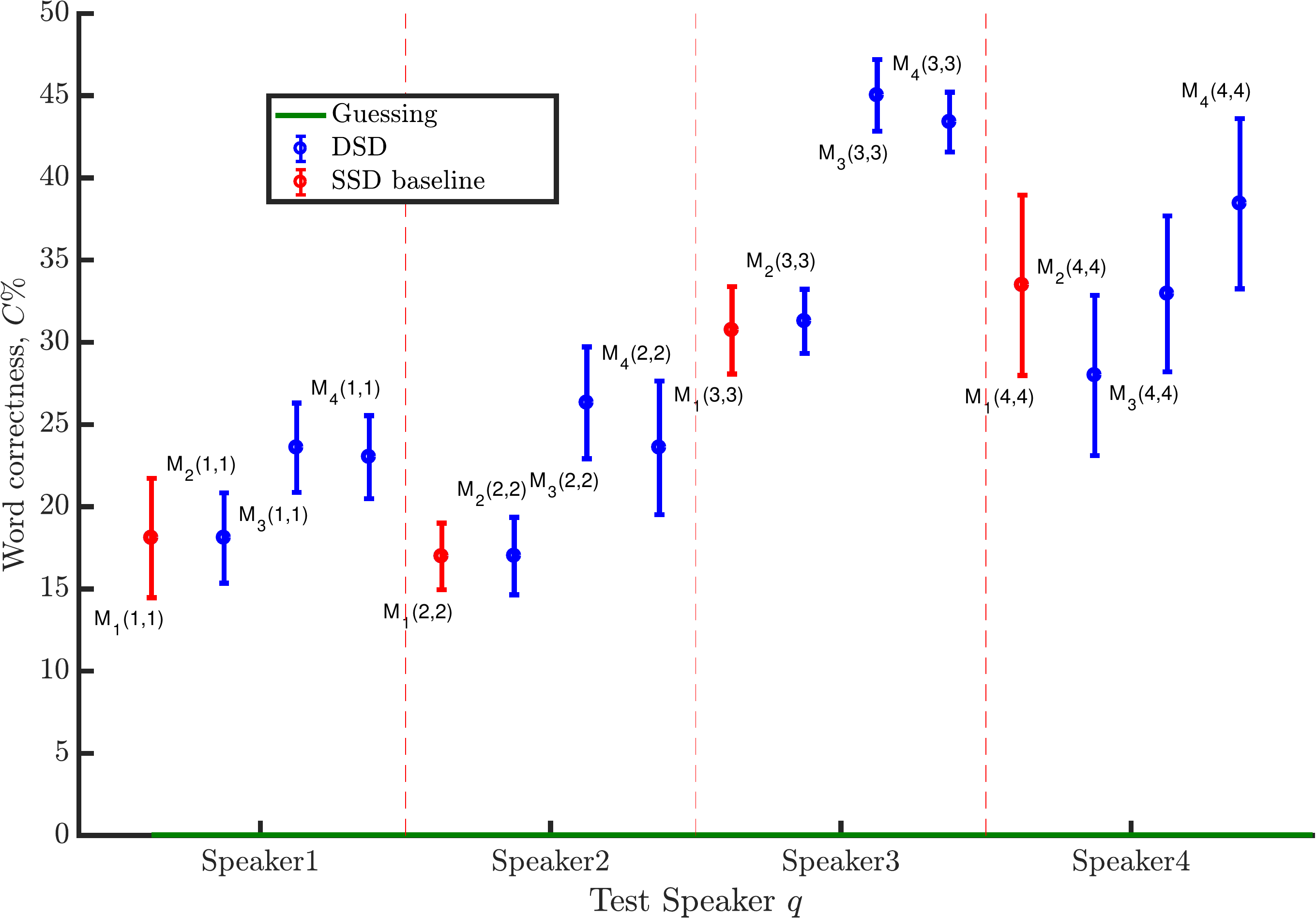} 
\caption{Word correctness, $C\pm1$s.e., of the AVL2 DSD tests. SSD baseline in red} 
\label{fig:correctnessDSD} 
\end{figure}

\begin{figure}[h] 
\centering 
\includegraphics[width=0.9\linewidth]{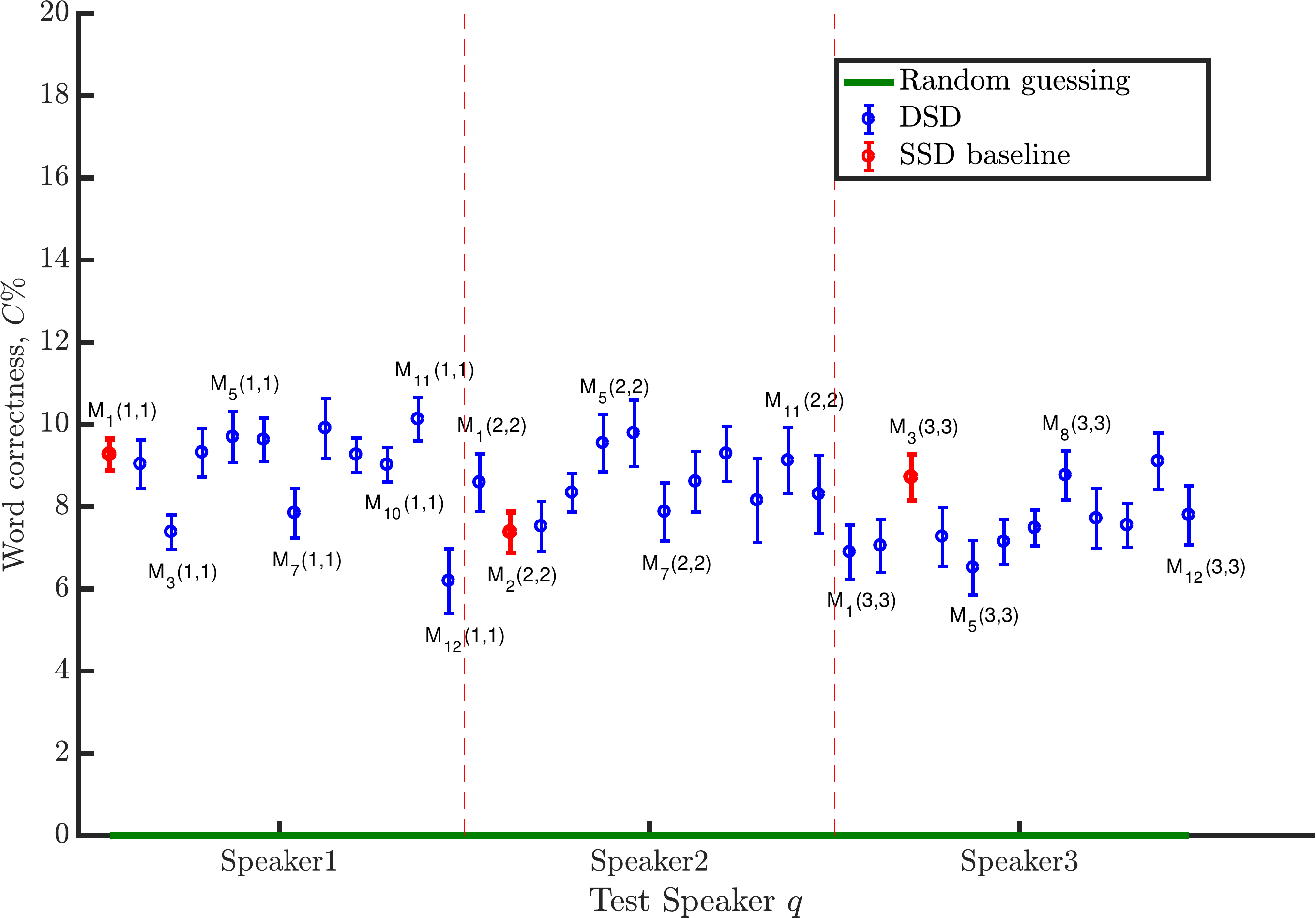} 
\caption{Word correctness, $C\pm1$s.e., of the RMAV speakers 1-3 DSD tests. SSD baseline in red} 
\label{fig:correctness1} 
\end{figure}

\begin{figure}[h] 
\centering 
\includegraphics[width=0.9\linewidth]{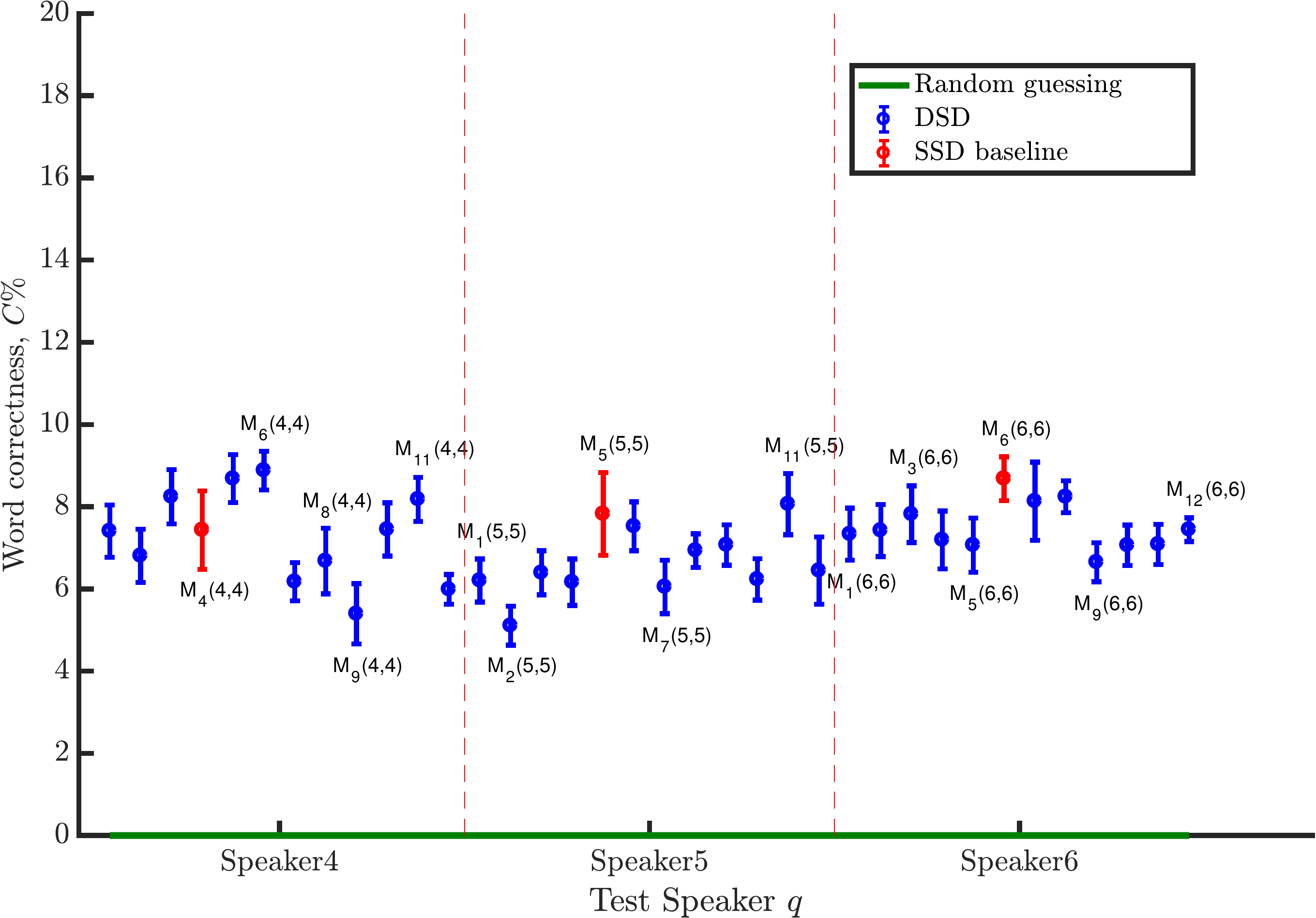} 
\caption{Word correctness, $C\pm1$s.e., of the RMAV speakers 4-6 DSD tests. SSD baseline in red} 
\label{fig:correctness2} 
\end{figure}

\begin{figure}[h] 
\centering 
\includegraphics[width=0.9\linewidth]{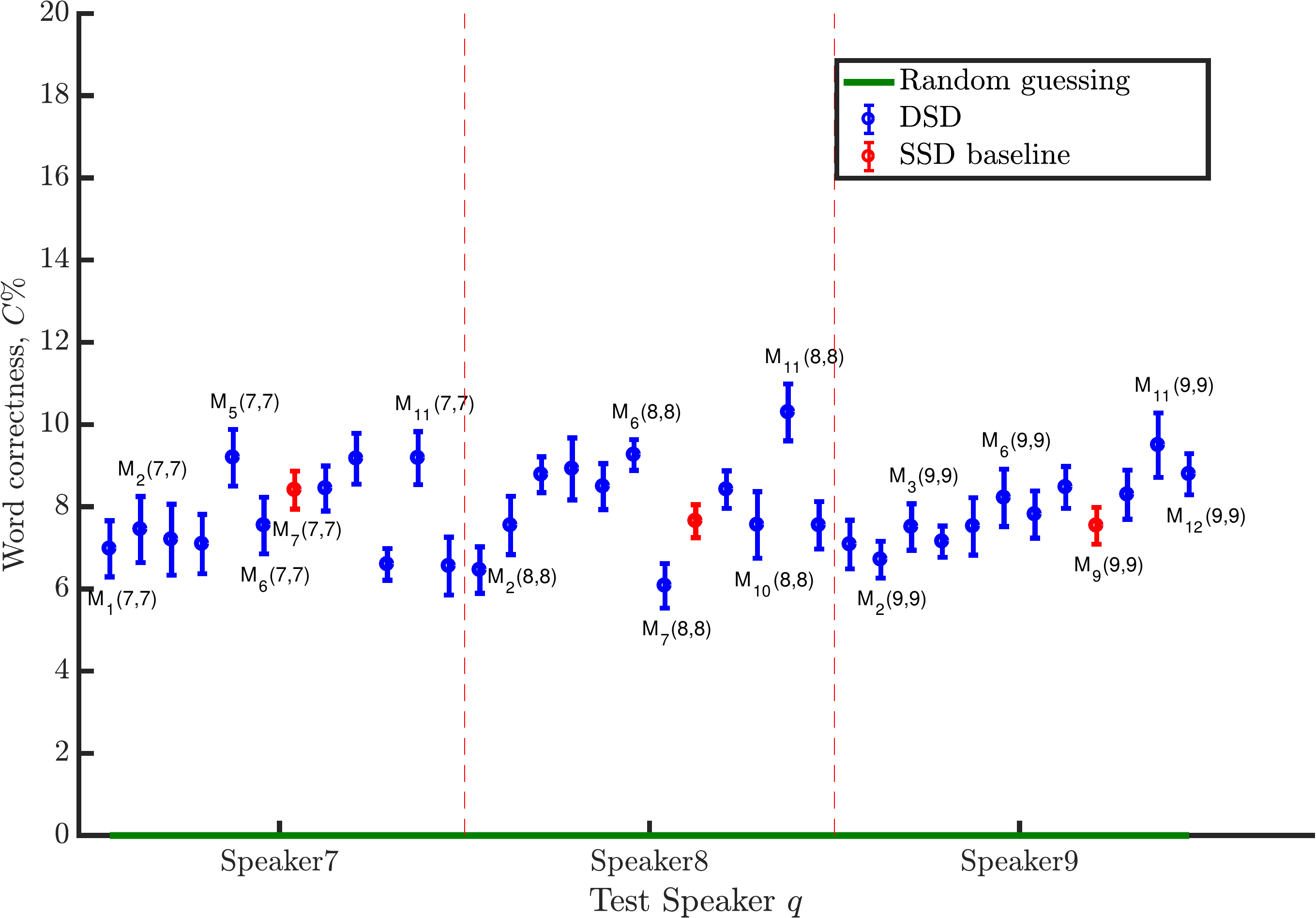} 
\caption{Word correctness, $C\pm1$s.e., of the RMAV speakers 7-9 DSD tests. SSD baseline in red} 
\label{fig:correctness3} 
\end{figure}

\begin{figure}[h] 
\centering 
\includegraphics[width=0.9\linewidth]{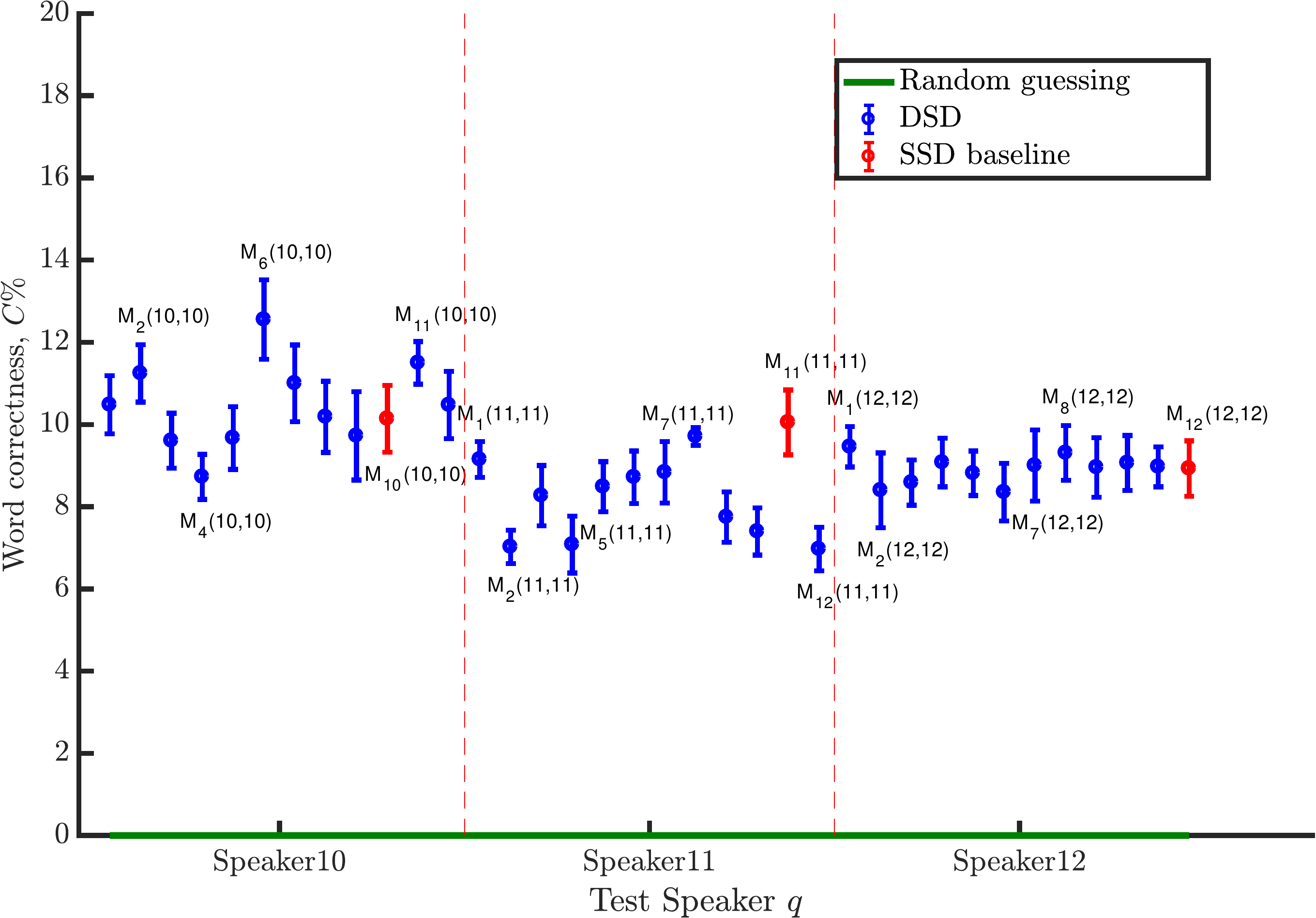} 
\caption{Word correctness, $C\pm1$s.e., of the RMAV speakers 10-12 DSD tests. SSD baseline in red} 
\label{fig:correctness4} 
\end{figure}

Figure~\ref{fig:correctnessDSD} shows the AVL2 DSD experiments from Table~\ref{tab:si}. In the DSD tests, the classifier is allowed to be trained on the relevant speaker, so the other tests are: $M_2(1,1)$, $M_3(1,1)$, $M_4(1,1)$ for Speaker 1, $M_1(2,2)$, $M_3(2,2)$, $M_4(2,2)$ for Speaker 2, $M_1(3,3)$, $M_2(3,3)$, $M_4(3,3)$ for Speaker 3 and finally $M_1(4,4)$, $M_2(4,4)$, $M_3(4,4)$ for Speaker 4. Now the word correctness has improved substantially which implies the previous poor performance in Figure~\ref{fig:indep_Corr} was not due to the choice of visemes but rather, the speaker-specific HMMs. The equivalent graphs for the $12$ RMAV speakers are in Figures~\ref{fig:correctness1},~\ref{fig:correctness2},~\ref{fig:correctness3} and~\ref{fig:correctness4}.

 
With continuous speech we can see the effects of unit selection. Using Speaker 1 for example, in Figure~\ref{fig:correctness1} the three maps $M_3$, $M_7$ and $M_{12}$ all significantly reduce the correctness for Speaker 1. In contrast, for Speaker 2 there are no significantly reducing maps but maps $M_1$, $M_4$, $M_5$, $M_6$, $M_9$, and $M_{11}$ all significantly improve the classification of Speaker 2. This suggests that it is not just the speakers' identity which is important for good classification but how it is used. Some individuals may simply be easier to lip read
or there are similarities between certain speakers which when learned on one speaker are able to better classify the visual distinctions between phonemes on similar other speakers.
 
In Figure~\ref{fig:correctness3} we see Speaker 7 is particularly robust to visual unit selection for the classifier labels. Conversely Speakers 5 (Figure~\ref{fig:correctness2}) and 12 (Figure~\ref{fig:correctness4}) are really affected by the visemes (or phoneme clusters). Its interesting to note this is a variability not previously considered, some speakers may be dependent on good visual classifiers and the mapping back to acoustics utterances, but others not so much. 
 
\begin{figure}[h] 
\centering 
\includegraphics[width=0.9\linewidth]{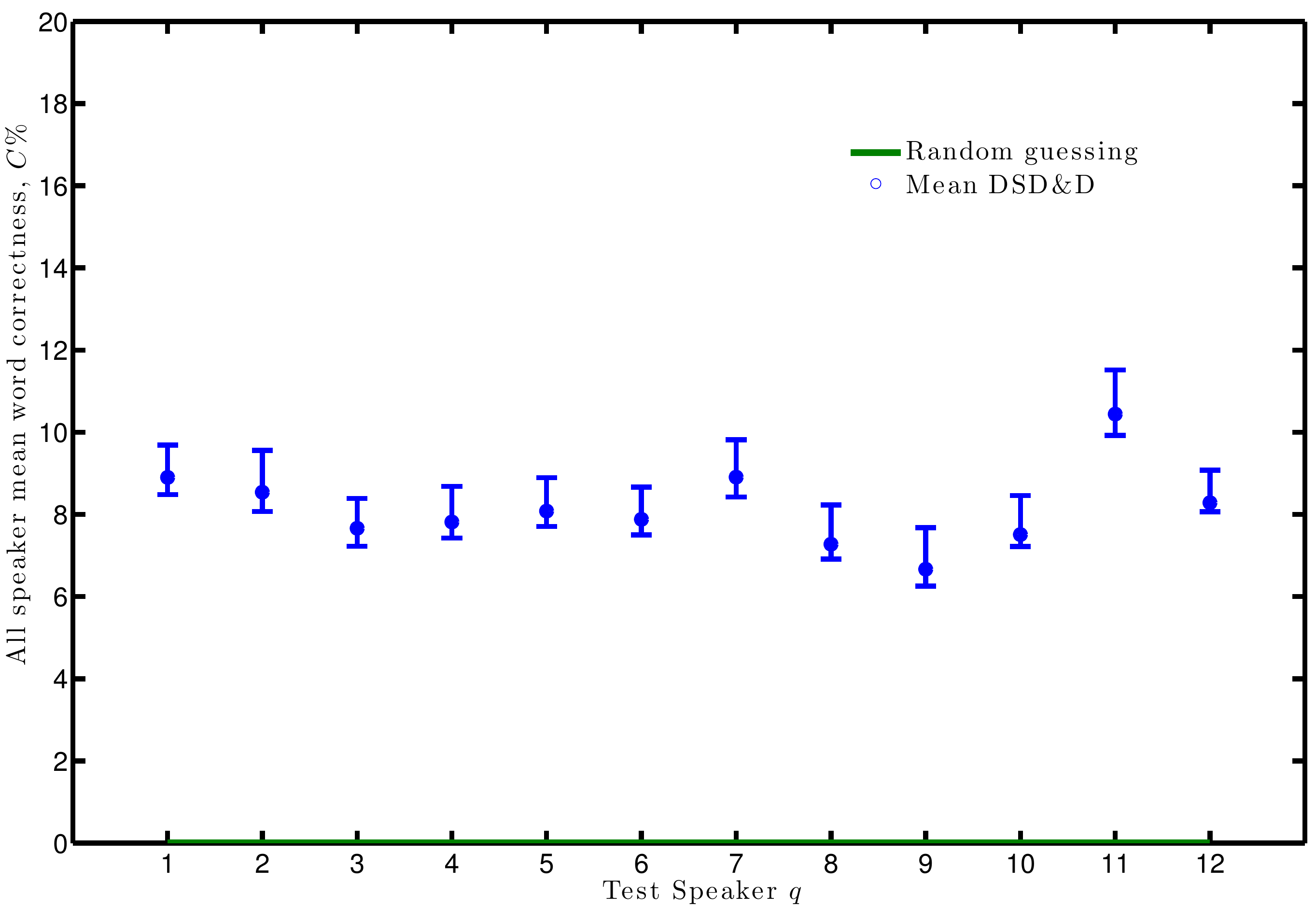} 
\caption{All-speaker mean word correctness, $C\pm1$standard error of the DSD tests} 
\label{fig:correctnessagg} 
\end{figure}

 Figure~\ref{fig:correctnessagg} shows the mean word correctness of the DSD classifiers per speaker in RMAV. The $y$-axis shows the \% word correctness and the $x$-axis is a speaker per point. We also plot random guessing and error bars of one standard error over the ten fold mean.
Speaker 11 is the best performing speaker irrespective of the P2V selected. All speakers have a similar standard error but a low mean within this bound. This suggests subject to speaker similarity, there is more possibility to improve classification correctness with another speakers visemes (if they include the original speakers visual cues) than to use weaker self-clustered visemes. 
 \FloatBarrier
\subsection{Weighting the $M_n$ effect on other speakers}
 To summarize the performance of DSD versus SSD we use scores. If DSD exceeds SSD by more than one standard error we score $+2$, or $-2$ if it is below. The scores $\pm1$ indicate differences within the standard error. The scores are shown in Tables~\ref{tab:weighting} and \ref{tab:weighting_lilir}.
\begin{table} [!ht]
\centering 
\caption{Weighted ranking scores from comparing the use of speaker-dependent maps for \emph{other} AVL2 speakers} 
\begin{tabular}{|l|r|r|r|r|} 
\hline 
& $M_1$ & $M_2$ & $M_3$ & $M_4$ \\ 
\hline \hline 
Sp01 & $0$ 		& $+1$		& $+2$		& $+2$ \\ 
Sp02 & $-1$		& $0$		& $+2$		& $+1$ \\ 
Sp03 & $-2$		& $-2$		& $0$		& $-1$ \\ 
Sp04 & $-1$		& $+1$		& $-1$		& $0$ \\ 
\hline
Total	 & $-4$		& $0$		& $\textbf{+3}$		& $\textbf{+2}$ \\ 
\hline 
\end{tabular} 
\label{tab:weighting} 
\end{table} 
$M_3$ scores the highest of the four AVL2 SSD maps, followed by $M_4$, $M_2$ and finally $M_1$ is the most susceptible to speaker identity in AVL2. It seems that the more similar to phoneme classes the visemes are, then the better the classification performance. This is consistent with Table~\ref{tab:homophones}, where the larger P2V maps create fewer homophones \cite{bear2015findingphonemes} 

\begin{table*} [!h]
\centering 
\caption{Weighted scores from comparing the use of speaker-dependent maps for \emph{other} speaker lipreading in continuous speech (RMAV speakers).} 
\begin{tabular}{|l|r|r|r|r|r|r|r|r|r|r|r|r|} 
\hline 
& $M_1$ & $M_2$ & $M_3$ & $M_4$ & $M_5$ & $M_6$ & $M_7$ & $M_8$ & $M_9$ & $M_{10}$ & $M_{11}$ & $M_{12}$ \\ 
 Num of visemes & 16&14&16&15&18&16&16&14&19&15&15&13\\
\hline \hline 
Sp01 & $0$ 		& $-1$		& $-2$		& $-2$		& $+1$		& $-1$		& $-1$		& $-1$		& $+1$		& $+1$		& $-1$		& $+1$ \\ 
Sp02 & $+2$ 		& $0$		& $+1$		& $+1$		& $+2$		& $+2$		& $+1$		& $+1$		& $+2$		& $+2$		& $+1$		& $+2$ \\ 
Sp03 & $-2$ 		& $-2$		& $0$		& $-2$		& $+1$		& $-1$		& $-1$		& $-2$		& $-2$		& $-2$		& $-2$		& $+1$ \\ 
Sp04 & $-2$ 		& $-1$		& $-1$		& $0$		& $+1$		& $+1$		& $-2$		& $-2$		& $+1$		& $-1$		& $-2$		& $+1$ \\ 
Sp05 & $-2$ 		& $-1$		& $+2$		& $-2$		& $0$		& $+1$		& $-1$		& $+2$		& $+1$		& $+2$		& $-1$		& $+2$ \\ 
Sp06 & $-1$ 		& $-1$		& $-1$		& $+1$		& $+2$		& $0$		& $+2$		& $-1$		& $-1$		& $+1$		& $+1$		& $+2$ \\ 
Sp07 & $+1$ 		& $-1$		& $-1$		& $+1$		& $+1$		& $+1$		& $0$		& $+1$		& $-1$		& $-1$		& $+1$		& $+1$ \\ 
Sp08 & $-1$ 		& $-1$		& $+1$		& $-1$		& $-1$		& $-2$		& $-2$		& $0$		& $+1$		& $+2$		& $+1$		& $+1$ \\ 
Sp09 & $-2$ 		& $-2$		& $-1$		& $-2$		& $-1$		& $-1$		& $-1$		& $-2$		& $0$		& $-1$		& $-2$		& $+1$ \\ 
Sp10 & $-2$ 		& $-2$		& $-1$		& $-1$		& $-1$		& $-2$		& $-2$		& $-2$		& $-2$		& $0$		& $-2$		& $-2$ \\ 
Sp11 & $-1$ 		& $+1$		& $-1$		& $+1$		& $+1$		& $-1$		& $+1$		& $-1$		& $-1$		& $+2$		& $0$		& $+2$ \\ 
Sp12 & $-1$ 		& $-2$		& $-2$		& $-1$		& $-1$		& $-2$		& $-2$		& $-2$		& $-2$		& $-1$		& $-2$		& $0$ \\ 
\hdashline
Total	 & $-9$		& $-11$		& $-6$		& $-7$ 		& $\textbf{+3}$ 		& $-5$ 		& $-8$ 		& $-9$ 		& $-3$ 		& $-4$ 		& $-8$ 		& $\textbf{+12}$ \\ 
\hline 
\end{tabular} 
\label{tab:weighting_lilir} 
\end{table*} 
In Table 7 of our supplementary material, we list the AVL2 speaker-dependent P2V maps. The phoneme pairs \{/\textschwa/, $/eh/$\}, \{$/m/$, $/n/$\} and \{$/ey/$, $/iy/$\} are present for three speakers and \{/\textturnv/, $/iy/$\} and \{$/l/$, $/m/$\} are pairs for two speakers. Of the single-phoneme visemes, \{/t\textipa{S}/\} is presented three times, \{$/f/$\}, \{$/k/$\}, \{$/w/$\} and \{$/z/$\} twice. We learn from Figure~\ref{fig:correctnessDSD} that the selection of incorrect units, whilst detrimental, is not as bad as training on alternative speakers. 

Table~\ref{tab:weighting_lilir} shows the scores for the 12 RMAV speakers. The speaker dependent map of Speaker 12 (right column) is one of only two ($M_{12}$ and $M_5$) which make an overall improvement on other speakers classification (they have positive values in the total row at the bottom of Table~\ref{tab:weighting_lilir}), and crucially, $M_{12}$ only has one speaker (Speaker 10) for whom the visemes in $M_{12}$ do not make an improvement in classification. The one other speaker P2V map which improves over other speakers is $M_5$. All others show a negative effect, this reinforces the observation that visual speech is dependent upon the individual but we also now have evidence there are exceptions to the rule. 
Table~\ref{tab:weighting_lilir} also lists the number of visemes within each set. All speaker-dependent sets are within the optimal range of $11$ to $35$ illustrated in \cite{bear2016decoding}. 
\FloatBarrier
\section{Speaker independence between sets of visemes} 
For isolated word classification the main conclusion of this section is shown by comparing Figures~\ref{fig:correctnessDSD} \&~\ref{fig:accuracy} with Figure~\ref{fig:indep_Corr}. The reduction in performance in Figure~\ref{fig:indep_Corr} is when the system classification models are trained on a speaker who is not the test speaker. This raised the question if this this degradation was due to the wrong choice of P2V map or speaker identity mismatch between the training and test data samples. We have concluded that, whilst the wrong unit labels are not conducive for good lipreading classification, is it not the choice of P2V map which causes significant degradation but rather the speaker identity. This regain of performance is irrespective of whether the map is chosen for a different speaker, multi-speaker or independently of the speaker. 
 
This observation is important as it tells us the repertoire of visual units across speakers does not vary significantly. This is comforting since the prospect of classification using a symbol alphabet which varies by speaker is daunting. 
There are differences between speakers, but not significant ones. However, we have seen some exceptions within the continuous speech speakers whereby the effect of the P2V map selection is more prominent and where sharing HMMs trained on non-test speakers has not been completely detrimental. This gives some hope with similar visual speakers, and with more `good' training data speaker independence, whether by classifier or viseme selection, might be possible.

To provide an analogy; in acoustic speech we could ask if an accented Norfolk speaker requires a different set of phonemes to a standard British speaker? The answer is no. They are represented by the same set of phonemes; but due to their individuality they use these phonemes in a different way.

Comparing the multi-speaker and SI maps, there are 11-12 visemes per set whereas in the single-speaker-dependent maps we have a range of 12 to 17. It is $M_3$ with 17 visemes, which out performs all other P2V maps. So we can conclude, there is a high risk of over-generalising a speaker-dependent P2V map when attempting multi-speaker or speaker-independent P2V mappings as we have seen with the RMAV experiments. 

Therefore we must consider it is not just the speaker-dependency which varies but also the contribution of each viseme within the set which also contributes to the word classification performance, an idea first shown in~\cite{bear2014some}. Here we have highlighted some phonemes which are a good subset of potentially independent visemes \{/\textschwa/, $/eh/$\}, \{$/m/$, $/n/$\} and \{$/ey/$, $/iy/$\}, and what these results present, is a combination of certain phoneme groups combined with some speaker-dependent visemes, where the latter provide a lower contribution to the overall classification would improve speaker-independent maps with speaker-dependent visual classifiers.

It is often said in machine lipreading there is high variability between speakers. This should now be clarified to state there is not a high variability of visual cues given a language, but there is high variability in trajectory between visual cues of an individual speakers with the same ground truth. In continuous speech we have seen how not just speaker identity affects the visemes (phoneme clusters) but also how the robustness of each speakers classification varies in response to changes in the viseme sets used. This implies a dependency upon the number of visemes within each set for individuals so this is what we investigate in the next section. 

Due to the many-to-one relationship in traditional mappings of phonemes to visemes, any resulting set of visemes will always be smaller than the set of phonemes. We know a benefit of this is more training samples per class which compensates for the limited data in currently available datasets but the disadvantage is generalization between different articulated sounds. To find an optimal set of viseme classes, we need to minimize the generalization to maintain good classification but also to maximize the training data available. 
\FloatBarrier
\section{Distance measurements between sets of heterogeneous visemes}
Our statistical measure is the Wilcoxon signed rank test \cite{wilcoxon1945individual}. Our intent is to move towards a distance measurement between the visual speech information for each speaker. We use a non-parametric method as we can not make assumptions about the distributions of the data, the individual P2V mappings re-distribute the data samples.

The signed rank test a non-parametric method which uses paired samples of values, to rank the population means of each pair-value. The sum of the signed ranks, $W$, is compared to the significance value. We use $\rho = 0.05$ for a $95\%$ confidence interval to determine significance, $p$. If $W<\rho$ then $p=1$ else $p=0$. 
The null hypothesis is there is no difference between the paired samples. In our case, this means that the speaker variation (represented in P2V maps) is not significant. In finding speakers who are significantly different, we hope to identify speakers who will be easier to adapt features between due to similarity in lip trajectory during speech. 

To compare the distances between the speaker-dependent P2V mappings, we use the Wilcoxon signed rank test which allows non-parametric pairwise comparison of speaker mean word correctness scores. Table~\ref{tab:corrAvl2} is the signed ranks $r$. Scores are underlined where the respective significance $\rho=1$. 
The respective continuous speech comparison is in Table~\ref{tab:corrRMAV}. 
Both tables are presented as a confusion matrix to compare all speakers with all others. The on-diagonal is always $r=1$ (in Tables~\ref{tab:corrAvl2} \& \ref{tab:corrRMAV}), 
This confirms speakers are identical when paired with themselves. 

\begin{table}[!ht]
\centering
\caption{Wilcoxon Signed Rank, $r$, for the AVL2 speakers}
\begin{tabular}{|l|rrrr|}
\hline
& Sp01 & Sp02 & Sp03 & Sp04 \\
\hline \hline
$M_1$ & 1.000 & 0.844 & \underline{0.016} & \underline{0.031}\\
$M_2$ & 0.844 & 1.000 & \underline{0.016} & \underline{0.016}\\
$M_3$ & \underline{0.016} & \underline{0.016} & 1.000 & 0.625\\
$M_4$ & \underline{0.031} & \underline{0.016} & 0.625 & 1.000\\
\hline
\end{tabular}
\label{tab:corrAvl2}
\end{table}

In Table~\ref{tab:corrAvl2}
we see an immediate split in the speakers. We can group speakers 1 and 2 together, and separately group speaker 3 with speaker 4. The similarity between speaker 1 and 2 ($r=0.844$) is greater than between speakers 3 and four ($r=0.625$). It is interesting that with a small dataset and a simple language model, there are clear distinctions between some speakers. 
\begin{table*}[!h]
\centering
\caption{Wilcoxon signed rank, $r$, for the RMAV speakers}
\begin{tabular}{|l|rrrrrrrrrrrr|}
\hline
& $M_1$ & $M_2$ & $M_3$ & $M_4$ & $M_5$ & $M_6$ & $M_7$ & $M_8$ & $M_9$ & $M_{10}$ & $M_{11}$ & $M_{12}$ \\
\hline \hline
$M_1$ & 1.000 & \underline{0.037} & 0.695 & 0.160 & 0.084 & \underline{0.020} & 0.275 & 0.193 & 0.193 & 0.375 & 0.508 & 0.275\\
$M_2$ & \underline{0.037} & 1.000 & 0.084 & \underline{0.037} & 1.000 & 0.922 & 0.084 & 1.000 & 0.625 & 0.064 & \underline{0.037} & \underline{0.020}\\
$M_3$ & 0.695 & 0.084 & 1.000 & 0.922 & 0.232 & 0.160 & 0.770 & 0.432 & 0.492 & 0.846 & 0.193 & 0.322\\
$M_4$ & 0.160 & \underline{0.037} & 0.922 & 1.000 & 0.322 & 0.232 & 0.492 & 0.432 & 0.334 & 0.922 & 0.105 & 0.105\\
$M_5$ & 0.084 & 1.000 & 0.232 & 0.322 & 1.000 & 1.000 & 0.275 & 1.000 & 1.000 & 0.131 & \underline{0.037} & 0.064\\
$M_6$ & \underline{0.020} & 0.922 & 0.160 & 0.232 & 1.000 & 1.000 & 0.193 & 1.000 & 1.000 & 0.152 & 0.064 & 0.064\\
$M_7$ & 0.275 & 0.084 & 0.770 & 0.492 & 0.275 & 0.193 & 1.000 & 0.275 & 0.375 & 0.770 & 0.375 & 0.232\\
$M_8$ & 0.193 & 1.000 & 0.432 & 0.432 & 1.000 & 1.000 & 0.275 & 1.000 & 0.922 & 0.232 & \underline{0.025} & 0.160\\
$M_9$ & 0.193 & 0.625 & 0.492 & 0.334 & 1.000 & 1.000 & 0.375 & 0.922 & 1.000 & 0.322 & 0.084 & 0.232\\
$M_{10}$ & 0.375 & 0.064 & 0.846 & 0.922 & 0.131 & 0.152 & 0.770 & 0.232 & 0.322 & 1.000 & 0.322 & 0.232\\
$M_{11}$ & 0.508 & \underline{0.037} & 0.193 & 0.105 & \underline{0.037} & 0.064 & 0.375 & \underline{0.025} & 0.084 & 0.322 & 1.000 & 0.770\\
$M_{12}$ & 0.275 & \underline{0.020} & 0.322 & 0.105 & 0.064 & 0.064 & 0.232 & 0.160 & 0.232 & 0.232 & 0.770 & 1.000\\
\hline
\end{tabular}
\label{tab:corrRMAV}
\end{table*}

Table~\ref{tab:corrRMAV} is the respective analysis for the RMAV speakers, these results are not clear cut. Four of the RMAV speakers are not significantly different from all others others, these are speakers 3, 7, 9, and 10. The significantly different speaker pairs are:
\begin{multicols}{3} 
\begin{itemize}
 \setlength\itemsep{0.001em} 
\item $M_1$, $M_{2}$ 
\item $M_1$, $M_{6}$ 
\item $M_2$, $M_{4}$ 
\item $M_2$, $M_{11}$ 
\item $M_2$, $M_{12}$ 
\item $M_5$, $M_{11}$ 
\item $M_{11}$, $M_{11}$ 
\end{itemize}
\end{multicols}

This observation reinforces the notion that some individual speakers have unique trajectories between visemes to make up their own visual speech signal, and idea first presented in \cite{bear2014some}, but here, others speakers (3, 7, 9, and 10) demonstrate a generalized pattern of visual speech units. 

We postulate that these four speakers could be more useful for speaker independent systems as generalizing from them is within a small data space. Also, adapting features between the other speakers would be more challenging as they have a greater distance between them. It is also possible that speaker adaptation may be complicated with our observation in section~\ref{sec:dsddResults}, that adaption between speakers could be directional. For example, if we look at speakers 1 and 2 from RMAV, we know they are significantly distinct (Table~\ref{tab:corrRMAV}) but, if we also reference the effect of the P2V maps of these speakers in Table~\ref{tab:weighting_lilir}, the visemes of speaker two insignificantly reduces the mean classification of speaker one whereas the visemes of speaker one significantly increases the mean classification of speaker two. This means that for this pair of speakers we prefer the visemes of speaker one. But this is not consistent for all significantly different visual speakers. Speaker pair 1 and 6 demonstrated both speakers classified more accurately with their own speaker-dependent visemes. This shows the complexity at the nub of speaker-independent lipreading systems for recognizing patterns of visual speech units, the units themselves are variable. 

\section{Conclusions}
By comparing Figure~\ref{fig:accuracy} with Figure~\ref{fig:indep_Corr} we show a substantial reduction in performance when the system is trained on non-test speakers. The question arises as to whether this degradation is due to the wrong choice of map or the wrong training data for the recognisers. We conclude that it is not the choice of map that causes degradation since we can retrain the HMMs and regain much of the performance. We regain performance irrespective of whether the map is chosen for a different speaker, multi-speaker or independently of the speaker. 

The sizes of the MS and SI maps built on continuous speech are fairly consistent, at most only $\pm2$ visemes per set. Whereas the SSD maps have a size range of six. We conclude there is high risk of over-generalizing a MS/SI P2V map. It is not only the speaker-dependency that varies but also the contribution of each viseme within the set which affects the word classification performance, an idea also shown in~\cite{bear2014some}. This suggests that a combination of certain MS visemes with some SD visemes would improve speaker-independent lipreading. We have shown exceptions where the P2V map choice is significant and where HMMs trained on non-test speakers has not been detrimental. This is evidence that with visually similar speakers, speaker-independent lipreading is probable. 
Furthermore, with continuous speech, we have shown that speaker dependent P2V maps significantly improve lipreading over isolated words. We attribute this to the co-articulation effects of visual speech on phoneme recognition confusions which in turn influences the speaker-dependent maps with linguistic or context information. This is supported by evidence from conventional lipreading systems which show the strength of language models in lipreading accuracy. 

We provide more evidence that speaker independence, even with unique trajectories between visemes for individual speakers, is likely to be achievable. What we need now is more understanding of the influence of language on visual gestures. What is in common, is the language between speakers. What we are seeking is an understanding of how language creates the gestures captured in visual speech features. 

We can address lipreading dependency on training speakers by generalizing to those speakers who are visually similar in viseme usage/trajectory through gestures. This is consistent with recent deep learning training methods. However here, we show that we should not need the big data volumes to do this generalization and presented evidence that adaptation between speakers may be directional meaning we can recognise speaker $A$ from speaker $B$ data, but not vice versa.

These are important conclusions because with the widespread adoption of deep learning and big data available, we trade-off data volumes and training time for improved accuracy. We have shown that if we can find a finite number of individuals whose visual speech gestures are similar enough to cover the whole test population, one could train on this much smaller data set for comparable results to lipreading big data. 

We have measured the distances/similarity between different speaker-dependent sets of visemes and shown there is minimal significant correlation supporting prior evidence about speaker heterogeneity in visual speech. However, these distances are variable and require further investigation. 

Our conclusion that it is the use, or trajectory of visemes, rather than the visemes themselves which vary by speaker suggests that there might be alternative approaches for finding phonemes in the visual speech channel of information. By this we mean that, using the linguistic premise that phonemes are consistent for all speakers, there could be a way of translating between strings of visemes which provide more information, thus are more discriminative for recognizing the phonemes actually spoken. This approach is consistent with deep learning methods which have excellent results when lipreading sentences rather than short units such as in \cite{AssaelSWF16}.

\section*{Acknowledgement}
 We gratefully acknowledge the assistance of Professor Stephen J Cox, formerly of the University of East Anglia, for his advice and guidance with this work. This work was conducted while Dr Bear was in receipt of a studentship from the UK Engineering and Physical Sciences Research Council (EPSRC).

\section*{References}
\bibliographystyle{elsarticle-num}
\bibliography{mybibfile}

\newpage
\appendix{Supplementary materials}
\section{Speaker-dependent phoneme-to-viseme maps}
\label{app:sd}
\begin{table}[!h] 
 \centering 
 	\caption{RMAV speakers 1 and 2} 
 	\begin{tabular} {| l | l || l | l |} 
 	\hline 
	\multicolumn{2}{| c ||}{Speaker 1 $M_1$} & \multicolumn{2}{ c |}{Speaker 2 $M_2$} \\ 
 	Viseme & Phonemes & Viseme & Phonemes \\ 
	\hline \hline 
/v01/ & /ae/ /ax/ /eh/ /\textrevepsilon/ /ey/ /\textsci/ & /v01/ & /\textschwa\textupsilon/ \\ 
 & /iy/ & & \\ 
/v02/ & /\textopeno/ /\textturnscripta/ /\textschwa\textupsilon/ & /v02/ & /ax/ /ay/ /eh/ /\textrevepsilon/ /\textsci/ /iy/ \\ 
 & & & /oh/ \\ 
 /v03/ & /\textsci\textschwa/ /\textopeno\textschwa/ & /v03/ & /\textturnv/ /\textipa{E}/ /ey/ \\ 
 /v04/ & /\textupsilon/ & /v04/ & /\textscripta\textupsilon/ /\textopeno\textschwa/ \\ 
 /v05/ & /\textturnv/ /\textipa{E}/ & /v05/ & /\textsci\textschwa/ \\ 
 /v06/ & /\textscripta/ /ay/ & /v06/ & /\textscripta/ /ae/ /\textopeno/ \\ 
 /v07/ & /ua/ & /v07/ & /\textupsilon/ \\ 
 /v08/ & /\textopeno\textsci/ & /v08/ & /ua/ \\ 
 /v09/ & /\textschwa/ & /v09/ & /\textopeno\textsci/ \\ 
 /v10/ & /\textscripta\textupsilon/ & /v10/ & /b/ /l/ /m/ /n/ /p/ /r/ \\ 
 & & & /s/ /\textipa{S}/ /t/ /v/ /w/ /z/ \\ 
 /v11/ & /d/ /\textipa{D}/ /f/ /d\textipa{Z}/ /k/ /l/ & /v11/ & /d/ /\textipa{D}/ /f/ /g/ /d\textipa{Z}/ /k/ \\ 
 & /m/ /n/ /p/ /s/ & & /ng/ \\ 
 /v12/ & /\textipa{N}/ /t/ /\textipa{T}/ /v/ /z/ & /v12/ & /hh/ /y/ \\ 
 /v13/ & /\textipa{S}/ & /v13/ & /t\textipa{S}/ /\textipa{T}/ \\ 
 /v14/ & /r/ /w/ /y/ & /v14/ & /\textipa{Z}/ \\ 
 /v15/ & /b/ /g/ /hh/ & /sil/ & /sil/ \\ 
 /v16/ & /t\textipa{S}/ & /sp/ & /sp/ \\ 
 /sil/ & /sil/ & /gar/ & /\textschwa/ /c/ \\ 
 /sp/ & /sp/ & & \\ 
 /gar/ & /\textipa{Z}/ /c/ & & \\ 
 \hline 
 	 \end{tabular} 
 \label{tab:lilirvamps1-2} 
 \end{table} 
 
\begin{table}[!h] 
 \centering 
 	\caption{RMAV speakers 3 and 4} 
 	\begin{tabular} {| l | l || l | l |} 
 	\hline
	\multicolumn{2}{| c ||}{Speaker 3 $M_3$} & \multicolumn{2}{ c |}{Speaker 4 $M_4$} \\ 
 	Viseme & Phonemes & Viseme & Phonemes \\ 
 	\hline \hline 
/v01/ & /\textopeno/ /ax/ /eh/ /\textrevepsilon/ /ey/ /\textsci/ & /v01/ & /\textschwa\textupsilon/ \\ 
 & /iy/ /oh/ /ow/ & & \\ 
/v02/ & /\textupsilon/ & /v02/ & /ae/ /\textopeno/ /ax/ /eh/ /\textrevepsilon/ /ey/ \\ 
 & & & /ih/ /iy/ /oh/ \\ 
 /v03/ & /ay/ /\textipa{E}/ /\textopeno\textschwa/ & /v03/ & /ay/ /\textsci\textschwa/ /\textopeno\textschwa/ \\ 
 /v04/ & /\textsci\textschwa/ & /v04/ & /\textturnv/ /\textscripta\textupsilon/ \\ 
 /v05/ & /ae/ /\textturnv/ & /v05/ & /\textscripta/ /\textipa{E}/ \\ 
 /v06/ & /\textsci\textschwa/ & /v06/ & /\textupsilon/ \\ 
 /v07/ & /\textscripta/ & /v07/ & /ua/ \\ 
 /v08/ & /ua/ & /v08/ & /k/ /l/ /m/ /n/ /p/ /r/ \\ 
 & & & /s/ /t/ /v/ /z/ \\ 
 /v09/ & /\textschwa/ & /v09/ & /d/ /\textipa{N}/ \\ 
 /v10/ & /\textscripta\textupsilon/ & /v10/ & /\textipa{D}/ /f/ /g/ /w/ \\ 
 /v11/ & /k/ /l/ /m/ /n/ /\textipa{N}/ /p/ & /v11/ & /d\textipa{Z}/ /\textipa{S}/ \\ 
 /v12/ & /f/ /r/ /s/ /\textipa{S}/ /t/ /\textipa{T}/ & /v12/ & /hh/ \\ 
 & /w/ /y/ /z/ & & \\ 
/v13/ & /t\textipa{S}/ /d/ /\textipa{D}/ /g/ & /v13/ & /t\textipa{S}/ /y/ \\ 
 /v14/ & /hh/ /d\textipa{Z}/ /v/ & /v14/ & /b/ /\textipa{T}/ \\ 
 /v15/ & /\textipa{Z}/ & /v15/ & /\textipa{Z}/ \\ 
 /v16/ & /b/ & /sil/ & /sil/ \\ 
 /sil/ & /sil/ & /sp/ & /sp/ \\ 
 /sp/ & /sp/ & /gar/ & /\textopeno\textsci/ /\textschwa/ /c/ \\ 
 /gar/ & /\textopeno\textsci/ /c/ & & \\ 
 \hline 
 	 \end{tabular} 
 \label{tab:lilirvamps3-4} 
 \end{table} 

\begin{table}[!h] 
 \centering 
 	\caption{RMAV speakers 5 and 6} 
 	\begin{tabular} {| l | l || l | l |} 
 	\hline 
	\multicolumn{2}{| c ||}{Speaker 5 $M_5$} & \multicolumn{2}{ c |}{Speaker 6 $M_6$} \\ 
 	Viseme & Phonemes & Viseme & Phonemes \\  
 	\hline \hline 
/v01/ & /\textschwa\textupsilon/ & /v01/ & /\textscripta/ /ae/ /\textopeno/ /ax/ /ay/ /ey/ \\ 
 & & & /ih/ /uw/ \\ 
 /v02/ & /\textopeno/ /ax/ /ay/ /eh/ /\textrevepsilon/ /ey/ & /v02/ & /iy/ /\textturnscripta/ /\textschwa\textupsilon/ \\ 
 & /ih/ /iy/ & & \\ 
/v03/ & /ae/ /\textopeno\textschwa/ & /v03/ & /\textrevepsilon/ \\ 
 /v04/ & /\textupsilon/ & /v04/ & /eh/ \\ 
 /v05/ & /\textscripta\textupsilon/ /ua/ & /v05/ & /\textipa{E}/ \\ 
 /v06/ & /\textturnv/ /\textturnscripta/ & /v06/ & /\textscripta\textupsilon/ \\ 
 /v07/ & /\textipa{E}/ & /v07/ & /\textturnv/ \\ 
 /v08/ & /\textscripta/ /\textsci\textschwa/ & /v08/ & /\textupsilon/ \\ 
 /v09/ & /\textschwa/ & /v09/ & /\textsci\textschwa/ \\ 
 /v10/ & /w/ & /v10/ & /\textschwa/ \\ 
 /v11/ & /y/ & /v11/ & /\textipa{D}/ /f/ /hh/ /l/ /m/ /\textipa{N}/ \\ 
 & & & /p/ /r/ /s/ /t/ \\ 
 /v12/ & /t/ /\textipa{T}/ /z/ & /v12/ & /\textipa{S}/ /v/ /y/ \\ 
 /v13/ & /l/ /m/ /n/ /p/ /r/ /s/ & /v13/ & /g/ /d\textipa{Z}/ /k/ /z/ \\ 
 & /\textipa{S}/ /v/ & & \\ 
/v14/ & /b/ /d\textipa{Z}/ & /v14/ & /b/ /d/ /w/ \\ 
 /v15/ & /g/ /hh/ & /v15/ & /t\textipa{S}/ /n/ \\ 
 /v16/ & /\textipa{D}/ /f/ /\textipa{N}/ & /v16/ & /\textipa{T}/ /\textipa{Z}/ \\ 
 /v17/ & /t\textipa{S}/ /d/ /k/ & /sil/ & /sil/ \\ 
 /v18/ & /\textipa{Z}/ & /sp/ & /sp/ \\ 
 /sil/ & /sil/ & /gar/ & /\textopeno\textsci/ /c/ /ua/ \\ 
 /sp/ & /sp/ & & \\ 
 /gar/ & /\textopeno\textsci/ /c/ & & \\ 
 \hline 
 	 \end{tabular} 
 \label{tab:lilirvamps5-6} 
 \end{table} 

\begin{table}[!h] 
 \centering 
 	\caption{RMAV speakers 7 and 8} 
 	\begin{tabular} {| l | l || l | l |} 
 	\hline 
	\multicolumn{2}{| c ||}{Speaker 7 $M_7$} & \multicolumn{2}{ c |}{Speaker 8 $M_8$} \\ 
 	Viseme & Phonemes & Viseme & Phonemes \\ 
 	\hline \hline 
/v01/ & /\textipa{E}/ /eh/ /\textrevepsilon/ & /v01/ & /ae/ /\textturnv/ /ay/ /eh/ /\textsci/ /iy/ \\ 
 & & & /ow/ /uh/ \\ 
 /v02/ & /ae/ /\textopeno/ /ax/ /ay/ /ey/ /\textsci/ & /v02/ & /\textopeno/ /ax/ /\textipa{E}/ /\textsci\textschwa/ \\ 
 & /iy/ /oh/ & & \\ 
/v03/ & /\textscripta/ /\textschwa\textupsilon/ /\textopeno\textschwa/ & /v03/ & /\textscripta/ /\textscripta\textupsilon/ /ey/ \\ 
 /v04/ & /\textupsilon/ & /v04/ & /ua/ /\textopeno\textschwa/ \\ 
 /v05/ & /ua/ & /v05/ & /\textrevepsilon/ /\textturnscripta/ \\ 
 /v06/ & /\textsci\textschwa/ & /v06/ & /\textschwa/ \\ 
 /v07/ & /\textscripta\textupsilon/ & /v07/ & /\textopeno\textsci/ \\ 
 /v08/ & /\textturnv/ & /v08/ & /b/ /d/ /\textipa{D}/ /f/ /k/ /l/ \\ 
 & & & /m/ /n/ /p/ /r/ /s/ /t/ \\ 
 /v09/ & /t\textipa{S}/ /d/ /\textipa{D}/ /g/ /k/ /l/ & /v09/ & /\textipa{S}/ /v/ /z/ \\ 
 & /m/ /n/ /p/ /r/ /t/ & & \\ 
/v10/ & /\textipa{S}/ & /v10/ & /d\textipa{Z}/ /w/ /y/ \\ 
 /v11/ & /s/ /v/ /w/ /y/ /z/ & /v11/ & /g/ \\ 
 /v12/ & /b/ /\textipa{N}/ /d\textipa{Z}/ & /v12/ & /hh/ /\textipa{T}/ \\ 
 /v13/ & /f/ /\textipa{T}/ & /v13/ & /t\textipa{S}/ /\textipa{N}/ \\ 
 /v14/ & /\textipa{N}/ & /v14/ & /\textipa{Z}/ \\ 
 /v15/ & /\textipa{Z}/ & /sil/ & /sil/ \\ 
 /v16/ & /hh/ & /sp/ & /sp/ \\ 
 /sil/ & /sil/ & /gar/ & /c/ \\ 
 /sp/ & /sp/ & & \\ 
 /gar/ & /\textopeno\textsci/ /c/ /\textschwa/ & & \\ 
 \hline 
 	 \end{tabular} 
 \label{tab:lilirvamps7-8} 
 \end{table} 

\begin{table}[!h] 
 \centering 
 	\caption{RMAV speakers 9 and 10} 
 	\begin{tabular} {| l | l || l | l |} 
 	\hline 
	\multicolumn{2}{| c ||}{Speaker 9 $M_9$} & \multicolumn{2}{ c |}{Speaker 10 $M_{10}$} \\ 
 	Viseme & Phonemes & Viseme & Phonemes \\ 
 	\hline \hline 
/v01/ & /ae/ /\textturnv/ /ey/ & /v01/ & /ax/ /ay/ /eh/ /ey/ /\textsci/ /iy/ \\ 
 & & & /oh/ /ow/ \\ 
 /v02/ & /\textopeno\textschwa/ & /v02/ & /\textscripta\textupsilon/ /\textopeno\textschwa/ \\ 
 /v03/ & /\textscripta/ /ax/ /ay/ /\textipa{E}/ /eh/ /\textrevepsilon/ & /v03/ & /\textupsilon/ \\ 
 & /ih/ /iy/ & & \\ 
/v04/ & /\textsci\textschwa/ /\textschwa\textupsilon/ & /v04/ & /ae/ /\textturnv/ /\textopeno/ /\textipa{E}/ \\ 
 /v05/ & /\textopeno/ /\textturnscripta/ & /v05/ & /\textrevepsilon/ /\textsci\textschwa/ \\ 
 /v06/ & /\textscripta\textupsilon/ & /v06/ & /\textscripta/ \\ 
 /v07/ & /\textupsilon/ & /v07/ & /ua/ \\ 
 /v08/ & /ua/ & /v08/ & /\textschwa/ \\ 
 /v09/ & /k/ /l/ /m/ /n/ /\textipa{N}/ /p/ & /v09/ & /k/ /l/ /m/ /n/ /p/ /r/ \\ 
 & /r/ /s/ /t/ /w/ & & /s/ /\textipa{S}/ /t/ /w/ /y/ /z/ \\ 
 /v10/ & /t\textipa{S}/ & /v10/ & /g/ /\textipa{T}/ /v/ \\ 
 /v11/ & /d/ /\textipa{D}/ /f/ /v/ & /v11/ & /t\textipa{S}/ /d/ /\textipa{D}/ /f/ /hh/ \\ 
 /v12/ & /\textipa{S}/ & /v12/ & /b/ \\ 
 /v13/ & /b/ /z/ & /v13/ & /\textipa{N}/ \\ 
 /v14/ & /\textipa{S}/ & /v14/ & /\textipa{Z}/ \\ 
 /v15/ & /hh/ & /v15/ & /d\textipa{Z}/ \\ 
 /v16/ & /y/ & /sil/ & /sil/ \\ 
 /v17/ & /g/ /d\textipa{Z}/ & /sp/ & /sp/ \\ 
 /v18/ & /\textipa{Z}/ & /gar/ & /\textopeno\textsci/ /c/ \\ 
 /v19/ & /\textipa{T}/ & & \\ 
 /sil/ & /sil/ & & \\ 
 /sp/ & /sp/ & & \\ 
 /gar/ & /\textopeno\textsci/ /\textschwa/ /c/ & & \\ 
 \hline 
 	 \end{tabular} 
 \label{tab:lilirvamps9-10} 
 \end{table} 

\begin{table}[!h] 
 \centering 
 	\caption{RMAV speakers 11 and 12} 
 	\begin{tabular} {| l | l || l | l |} 
 	\hline 
	\multicolumn{2}{| c ||}{Speaker 11 $M_{11}$} & \multicolumn{2}{ c |}{Speaker 12 $M_{12}$} \\ 
 	Viseme & Phonemes & Viseme & Phonemes \\  
 	\hline \hline 
/v01/ & /\textipa{E}/ /\textupsilon/ & /v01/ & /ax/ /ay/ /eh/ /ey/ /\textsci/ /iy/ \\ 
 & & & /ow/ /uw/ \\ 
 /v02/ & /ae/ /eh/ /\textrevepsilon/ /ey/ /\textsci/ /iy/ & /v02/ & /\textscripta/ /ae/ /\textturnv/ /\textopeno/ /\textturnscripta/ \\ 
 & /oh/ & & \\ 
/v03/ & /\textturnv/ /\textopeno/ /ax/ & /v03/ & /\textipa{E}/ /\textsci\textschwa/ /\textopeno\textsci/ \\ 
 /v04/ & /ay/ /\textopeno\textschwa/ & /v04/ & /ua/ \\ 
 /v05/ & /\textsci\textschwa/ /\textschwa\textupsilon/ & /v05/ & /\textupsilon/ \\ 
 /v06/ & /\textscripta\textupsilon/ & /v06/ & /\textrevepsilon/ \\ 
 /v07/ & /\textscripta/ & /v07/ & /\textscripta\textupsilon/ \\ 
 /v08/ & /k/ /l/ /n/ /\textipa{N}/ /p/ /r/ & /v08/ & /w/ \\ 
 & /s/ /\textipa{S}/ /z/ & & \\ 
/v09/ & /m/ /t/ /\textipa{T}/ /v/ & /v09/ & /k/ /l/ /m/ /n/ /p/ /r/ \\ 
 & & & /s/ /\textipa{S}/ /t/ /th/ \\ 
 /v10/ & /g/ & /v10/ & /v/ /\textipa{Z}/ /t\textipa{S}/ \\ 
 /v11/ & /w/ & /v11/ & /y/ /b/ \\ 
 /v12/ & /t\textipa{S}/ /d\textipa{Z}/ & /v12/ & /d/ /\textipa{D}/ /f/ /g/ /n/ /\textipa{N}/ \\ 
 /v13/ & /b/ /d/ /\textipa{D}/ /f/ & /v13/ & /hh/ /d\textipa{Z}/ /z/ \\ 
 /v14/ & /hh/ /y/ & /sil/ & /sil/ \\ 
 /v15/ & /\textipa{Z}/ & /sp/ & /sp/ \\ 
 /sil/ & /sil/ & /gar/ & /c/ /\textschwa/ \\ 
 /sp/ & /sp/ & & \\ 
 /gar/ & /\textopeno\textsci/ /ua/ /c/ /\textschwa/ & & \\ 
 \hline 
 	 \end{tabular} 
 \label{tab:lilirvamps11-12} 
 \end{table} 
 
 \begin{table}[!h] 
\centering 
\caption{AVL2 speakers 1 to 4} 
\begin{tabular} {| l | l || l | l |} 
\hline 
\multicolumn{2}{| c ||}{Speaker 1 $M_1$} & \multicolumn{2}{ c |}{Speaker 2 $M_2$} \\ 
Viseme & Phonemes & Viseme & Phonemes \\ 
\hline \hline 
/v01/	& /\textturnv/ /iy/ /\textschwa\textupsilon/ /uw/ 	& /v01/	& /ay/ /ey/ /iy/ /uw/		\\ 
/v02/	& /\textschwa/ /eh/ /ey/		& /v02/	& /\textschwa\textupsilon/			\\ 
/v03/	& /\textscripta/ /ay/			& /v03/	& /\textschwa/			\\ 
/v04/	& /d/ /s/ /t/					& /v04/	& /eh/			\\ 
/v05/	& /t\textipa{S}/ /l/ 			& /v05/	& /\textturnv/ 			\\ 
/v06/	& /m/ /n/					& /v06/ 	& /\textschwa/			\\ 
/v07/	& /d\textipa{Z}/ /v/			& /v07/	& /d\textipa{Z}/ /p/ /y/ 		\\ 
/v08/	& /b/ /y/ 					& /v08/	& /l/ /m/ /n/ 		\\ 
/v09/	& /k/						& /v09/	& /v/ /w/			\\ 
/v10/	& /z/ 						& /v10/	& /d/ /b/			\\ 
/v11/	& /w/						& /v11/	& /f/ /s/			\\ 
/v12/	& /f/						& /v12/	& /t/ 				\\ 
&							& /v13/	& /k/				\\ 
&							& /v14/	& /t\textipa{S}/ 			\\ 
/sil/ 	& /sil/					& /sil/	& /sil/			\\ 
/garb/& /\textipa{E}/ /\textturnscripta/ /\textopeno/ /r/ /p/	& /garb/	& /\textipa{E}/ /\textturnscripta/ /\textopeno/ /r/ /z/	\\ 
\hline \hline 
\multicolumn{2}{| c ||}{Speaker 3 $M_3$} & \multicolumn{2}{ c |}{Speaker 4 $M_4$} \\ 
Viseme & Phonemes			& Viseme & Phonemes \\ 
\hline \hline 
/v01/	& /ey/ /iy/			& /v01/	& /\textturnv/ /ay/ /ey/ /iy/	\\ 
/v02/	& /\textschwa/ /eh/			& /v02/	& /\textschwa/ /eh/			\\ 
/v03/	& /ay/			& /v03/	& /\textschwa/			\\ 
/v04/	& /\textturnv/			& /v04/	& /\textschwa\textupsilon/			\\ 
/v05/	& /\textschwa/			& /v05/	& /uw/			\\ 
/v06/	& /\textschwa\textupsilon/			& /v06/	& /m/ /n/			\\ 
/v07/	& /uw/			& /v07/	& /k/ /l/			\\ 
/v08/	& /d/ /p/ /t/			& /v08/	& /d\textipa{Z}/ /t/			\\ 
/v09/	& /l/ /m/			& /v09/	& /d/ /s/			\\ 
/v10/	& /k/ /w/			& /v10/	& /w/				\\ 
/v11/	& /f/ /n/			& /v11/	& /f/				\\ 
/v12/	& /b/ /s/			& /v12/	& /v/				\\ 
/v13/	& /v/				& /v13/	& /t\textipa{S}/ 		\\ 
/v14/	& /d\textipa{Z}/ 		& /v14/	& /b/				\\ 
/v15/	& /t\textipa{S}/ 		& /v15/	& /y/				\\ 
/v16/	& /y/				&		&				\\ 
/v17/	& /z/				& 		&				\\ 
/sil/		& /sil/			& /sil/	& /sil/			\\ 
/garb/	& /\textipa{E}/ /\textturnscripta/ /\textopeno/ /r/	& /garb/	& /\textipa{E}/ /\textturnscripta/ /\textopeno/ /r/ /p/ /z/ \\ 
\hline 
\end{tabular} 
\label{tab:td_v} 
\end{table}
\vfill\newpage
 \section*{Multi-speaker phoneme-to-viseme maps}
 \label{app:msmaps}
 \begin{table}[!h] 
\centering 
\caption{AVL2 speakers (left) and RMAV speakers (right)} 
\begin{tabular}{| l | l || l | l |} 
\hline
\multicolumn{2}{|c||}{AVL2} & \multicolumn{2}{|c|}{RMAV} \\
Viseme & Phonemes & Viseme & Phonemes \\ 
\hline \hline 
/v01/ & /\textturnv/ /ay/ /ey/ /iy/ & /v01/ & /\textscripta/ /\ae/ /\textturnv/ /\textopeno/ /\textschwa/ /ay/ /\textipa{E}/ /eh/ \\
& /\textschwa\textupsilon/ /uw/ & & /\textrevepsilon/ /ey/ /\textsci \textschwa/ /\textsci/ /iy/ /\textturnscripta/ /\textschwa \textupsilon/  \\ 
/v02/ & /\textschwa/ /eh/ & /v02/ & /\textopeno\textschwa/ /\textupsilon/ /\textopeno\textschwa/ \\ 
/v03/ & /\textscripta/ & /v03/ & /\textscripta\textupsilon/ \\ 
/v04/ & /d/ /s/ /t/ /v/ & /v04/ & /\textopeno\textsci/ \\ 
/v05/ & /f/ /l/ /n/ & /v05/ & /\textschwa/ \\ 
/v06/ & /b/ /w/ /y/ & /v06/ & /b/ /t\textipa{S}/ /d/ /\textipa{D}/ /f/ /g/ /\textipa{H}/ /d\textipa{Z}/ \\
& & &  /k/ /l/ /m/ /n/ /\textipa{N}/ /p/ /r/ /s/ \\
& & & /\textipa{S}/ /t/ /\textipa{T}/ /v/ /w/ /y/ /z/  \\ 
/v07/ & /d\textipa{Z}/ & /sil/ & /sil/ \\ 
/v08/ & /z/ & /sp/ & /sp/ \\ 
/v09/ & /p/ & /gar/ & /\textipa{Z}/ /c/ \\ 
/v10/ & /m/ & & \\ 
/v11/ & /k/ & & \\ 
/v12/ & /t\textipa{S}/ & & \\ 
/sil/ & /sil/ & & \\ 
/gar/ & /\textipa{E}/ /\textturnscripta/ /\textopeno/ /r/ & & \\	\hline 
\end{tabular} 
\label{tab:mt_v} 
\end{table} 
\vfill\newpage
\section*{Speaker - Independent (SI) phoneme-to-viseme maps}
\label{app:simaps}
\begin{table}[!h] 
 \centering 
 	\caption{RMAV speakers 1 and 2} 
 	\begin{tabular} {| l | l || l | l |} 
 	\hline 
	\multicolumn{2}{| c ||}{Speaker 1 $M_{!1}$} & \multicolumn{2}{ c |}{Speaker 2 $M_{!2}$} \\ 
 	Viseme & Phonemes			& Viseme & Phonemes \\ 
 	\hline \hline 
/v01/ & /\textscripta/ /ae/ /\textturnv/ /\textopeno/ /ax/ /ay/ & /v01/ & /\textscripta/ /ae/ /\textturnv/ /\textopeno/ /ax/ /ay/ \\ 
 & /\textipa{E}/ /eh/ /\textrevepsilon/ /ey/ /\textsci\textschwa/ /\textsci/ && /\textipa{E}/ /eh/ /\textrevepsilon/ /ey/ /\textsci\textschwa/ /\textsci/ \\
& /iy/ /\textturnscripta/ /\textschwa\textupsilon/ & & /iy/ /\textturnscripta/ /\textschwa\textupsilon/ \\ 
 /v02/ & /ua/ /\textupsilon/ /\textopeno\textschwa/ & /v02/ & /ua/ /\textupsilon/ /\textopeno\textschwa/ \\ 
 /v03/ & /\textscripta\textupsilon/ & /v03/ & /\textscripta\textupsilon/ \\ 
 /v04/ & /\textopeno\textsci/ & /v04/ & /\textopeno\textsci/ \\ 
 /v05/ & /\textschwa/ & /v05/ & /\textschwa/ \\ 
 /v06/ & /b/ /t\textipa{S}/ /d/ /\textipa{D}/ /f/ /g/ & /v06/ & /b/ /t\textipa{S}/ /d/ /\textipa{D}/ /f/ /g/ \\ 
 & /hh/ /d\textipa{Z}/ /k/ /l/ /m/ & & /hh/ /d\textipa{Z}/ /k/ /l/ /m/ \\
 & /n/ /\textipa{N}/ /p/ /r/ /s/ /\textipa{S}/ && /n/ /\textipa{N}/ /p/ /r/ /s/ /\textipa{S}/ \\
 & /t/ /\textipa{T}/ /v/ /w/ /y/ /z/ & &/t/ /\textipa{T}/ /v/ /w/ /y/ /z/ \\ 
 /v07/ & /\textipa{Z}/ & /v07/ & /\textipa{Z}/ \\ 
 /sil/ & /sil/ & /sil/ & /sil/ \\ 
 /sp/ & /sp/ & /sp/ & /sp/ \\ 
 /gar/ & /c/ & /gar/ & /c/ \\ 
 \hline 
 	 \end{tabular} 
 \label{tab:lilirsimaps1-2} 
 \end{table} 
 
 \begin{table}[!h] 
 \centering 
 	\caption{RMAV speakers 3 and 4} 
 	\begin{tabular} {| l | l || l | l |} 
 	\hline \multicolumn{2}{| c ||}{Speaker 3 $M_{!3}$} & \multicolumn{2}{ c |}{Speaker 4 $M_{!4}$} \\ 
 	Viseme & Phonemes			& Viseme & Phonemes \\ 
 	\hline \hline 
/v01/ & /\textscripta/ /ae/ /\textturnv/ /\textopeno/ /ax/ /ay/ & /v01/ & /\textscripta/ /ae/ /\textturnv/ /\textopeno/ /ax/ /ay/ \\ 
 & /\textipa{E}/ /eh/ /\textrevepsilon/ /ey/ /\textsci\textschwa/ /\textsci/ & & /\textipa{E}/ /eh/ /\textrevepsilon/ /ey/ /\textsci\textschwa/ /\textsci/ \\
 & /iy/ /\textturnscripta/ /\textschwa\textupsilon/ & &/iy/ /\textturnscripta/ /\textschwa\textupsilon/ \\ 
 /v02/ & /ua/ /\textupsilon/ /\textopeno\textschwa/ & /v02/ & /ua/ /\textupsilon/ /\textopeno\textschwa/ \\ 
 /v03/ & /\textscripta\textupsilon/ /\textopeno\textsci/ & /v03/ & /\textscripta\textupsilon/ \\ 
 /v04/ & /\textschwa/ & /v04/ & /\textopeno\textsci/ \\ 
 /v05/ & /b/ /t\textipa{S}/ /d/ /\textipa{D}/ /f/ /g/ & /v05/ & /\textschwa/ \\ 
 & /hh/ /d\textipa{Z}/ /k/ /l/ /m/ /n/ & & \\
 & /\textipa{N}/ /p/ /r/ /s/ /\textipa{S}/ /t/ & & \\
 & /\textipa{T}/ /v/ /w/ /y/ /z/ & & \\ 
/v06/ & /\textipa{Z}/ & /v06/ & /b/ /t\textipa{S}/ /d/ /\textipa{D}/ /f/ /g/ \\ 
 & & & /hh/ /d\textipa{Z}/ /k/ /l/ /m/ \\ 
 & & & /n/ /\textipa{N}/ /p/ /r/ /s/ /\textipa{S}/ /t/ \\
 && &/\textipa{T}/ /v/ /w/ /y/ /z/ \\ 
 /sil/ & /sil/ & /v07/ & /\textipa{Z}/ \\ 
 /sp/ & /sp/ & /sil/ & /sil/ \\ 
 /gar/ & /c/ & /sp/ & /sp/ \\ 
 & & /gar/ & /c/ \\ 
 \hline 
 	 \end{tabular} 
 \label{tab:lilirsimaps3-4} 
 \end{table} 
 
\begin{table}[!h] 
 \centering 
 	\caption{RMAV speakers 5 and 6} 
 	\begin{tabular} {| l | l || l | l |} 
 	\hline 
	\multicolumn{2}{| c ||}{Speaker 5 $M_{!5}$} & \multicolumn{2}{ c |}{Speaker 6 $M_{!6}$} \\ 
 	Viseme & Phonemes			& Viseme & Phonemes \\ 
	 	\hline \hline 
/v01/ & /\textscripta/ /ae/ /\textturnv/ /\textopeno/ /ax/ /ay/ & /v01/ & /\textscripta/ /ae/ /\textturnv/ /\textopeno/ /ax/ /ay/ \\ 
 & /\textipa{E}/ /eh/ /\textrevepsilon/ /ey/ /\textsci\textschwa/ /\textsci/ && /\textipa{E}/ /eh/ /\textrevepsilon/ /ey/ /\textsci\textschwa/ /\textsci/ \\
 & /iy/ /\textturnscripta/ /\textschwa\textupsilon/ & & /iy/ /\textturnscripta/ /\textschwa\textupsilon/ \\ 
 /v02/ & /ua/ /\textupsilon/ /\textopeno\textschwa/ & /v02/ & /ua/ /\textupsilon/ /\textopeno\textschwa/ \\ 
 /v03/ & /\textscripta\textupsilon/ /\textopeno\textsci/ & /v03/ & /\textscripta\textupsilon/ \\ 
 /v04/ & /\textschwa/ & /v04/ & /\textopeno\textsci/ \\ 
 /v05/ & /b/ /t\textipa{S}/ /d/ /\textipa{D}/ /f/ /g/ & /v05/ & /\textschwa/ \\ 
 & /hh/ /d\textipa{Z}/ /k/ /l/ /m/ /n/ & & \\
 & /\textipa{N}/ /p/ /r/ /s/ /\textipa{S}/ /t/ & & \\
 & /\textipa{T}/ /v/ /w/ /y/ /z/ & & \\ 
/v06/ & /\textipa{Z}/ & /v06/ & /b/ /t\textipa{S}/ /d/ /\textipa{D}/ /f/ /g/ \\ 
& & & /hh/ /d\textipa{Z}/ /k/ /l/ /m/ /n/ \\
& & & /\textipa{N}/ /p/ /r/ /s/ /\textipa{S}/ /t/ \\
& & & /\textipa{T}/ /v/ /w/ /y/ /z/ \\ 
 /sil/ & /sil/ & /v07/ & /\textipa{Z}/ \\ 
 /sp/ & /sp/ & /sil/ & /sil/ \\ 
 /gar/ & /c/ & /sp/ & /sp/ \\ 
 & & /gar/ & /c/ \\ 
 \hline 
 	 \end{tabular} 
 \label{tab:lilirsimaps5-6} 
 \end{table} 
 
\begin{table}[!h] 
 \centering 
 	\caption{RMAV speakers 7 and 8} 
 	\begin{tabular} {| l | l || l | l |} 
 	\hline 
	\multicolumn{2}{| c ||}{Speaker 7 $M_{!7}$} & \multicolumn{2}{ c |}{Speaker 8 $M_{!8}$} \\ 
 	Viseme & Phonemes			& Viseme & Phonemes \\ 
 	\hline \hline 
/v01/ & /\textscripta/ /ae/ /\textturnv/ /\textopeno/ /ax/ /ay/ & /v01/ & /\textscripta/ /ae/ /\textturnv/ /\textopeno/ /ax/ /ay/ \\ 
 & /\textipa{E}/ /eh/ /\textrevepsilon/ /ey/ /\textsci\textschwa/ /\textsci/ & & /\textipa{E}/ /eh/ /\textrevepsilon/ /ey/ /\textsci\textschwa/ /\textsci/ \\\
 & /iy/ /\textturnscripta/ /\textschwa\textupsilon/ & & /iy/ /\textturnscripta/ /\textschwa\textupsilon/ \\ 
 /v02/ & /ua/ /\textupsilon/ /\textopeno\textschwa/ & /v02/ & /\textupsilon/ /\textopeno\textschwa/ \\ 
 /v03/ & /\textscripta\textupsilon/ & /v03/ & /ua/ \\ 
 /v04/ & /\textopeno\textsci/ & /v04/ & /\textschwa/ \\ 
 /v05/ & /\textschwa/ & /v05/ & /\textscripta\textupsilon/ /\textopeno\textsci/ \\ 
 /v06/ & /b/ /t\textipa{S}/ /d/ /\textipa{D}/ /f/ /g/ & /v06/ & /b/ /t\textipa{S}/ /d/ /\textipa{D}/ /f/ /g/ \\ 
 & /hh/ /d\textipa{Z}/ /k/ /l/ /m/ /n/ && /hh/ /d\textipa{Z}/ /k/ /l/ /m/ /n/ \\
 & /\textipa{N}/ /p/ /r/ /s/ /\textipa{S}/ /t/ & & /\textipa{N}/ /p/ /r/ /s/ /\textipa{S}/ /t/ \\
 & /\textipa{T}/ /v/ /w/ /y/ /z/ & & /\textipa{T}/ /v/ /w/ /y/ /z/ \\ 
 /v07/ & /\textipa{Z}/ & /v07/ & /\textipa{Z}/ \\ 
 /sil/ & /sil/ & /sil/ & /sil/ \\ 
 /sp/ & /sp/ & /sp/ & /sp/ \\ 
 /gar/ & /c/ & /gar/ & /c/ \\ 
 \hline 
 	 \end{tabular} 
 \label{tab:lilirsimaps7-8} 
 \end{table} 
 
\begin{table}[!h] 
 \centering 
 	\caption{RMAV speakers 9 and 10} 
 	\begin{tabular} {| l | l || l | l |} 
 	\hline 
	\multicolumn{2}{| c ||}{Speaker 9 $M_{!9}$} & \multicolumn{2}{ c |}{Speaker 10 $M_{!10}$} \\ 
 	Viseme & Phonemes			& Viseme & Phonemes \\ 
 	\hline \hline 
/v01/ & /\textscripta/ /ae/ /\textturnv/ /\textopeno/ /ax/ /ay/ & /v01/ & /\textupsilon/ \\ 
 & /\textipa{E}/ /eh/ /\textrevepsilon/ /ey/ /\textsci\textschwa/ /\textsci/ & & \\
 & /iy/ /\textturnscripta/ /\textschwa\textupsilon/ & & \\ 
/v02/ & /ua/ /\textupsilon/ /\textopeno\textschwa/ & /v02/ & /\textscripta/ /ae/ /\textturnv/ /\textopeno/ /ax/ /ay/ \\ 
 & & & /\textipa{E}/ /eh/ /\textrevepsilon/ /ey/ /\textsci\textschwa/ /\textsci/ \\
 & & & /iy/ /\textturnscripta/ /\textschwa\textupsilon/ \\ 
 /v03/ & /\textscripta\textupsilon/ & /v03/ & /\textscripta\textupsilon/ /ua/ /\textopeno\textschwa/ \\ 
 /v04/ & /\textopeno\textsci/ & /v04/ & /\textopeno\textsci/ \\ 
 /v05/ & /\textschwa/ & /v05/ & /\textschwa/ \\ 
 /v06/ & /b/ /t\textipa{S}/ /d/ /\textipa{D}/ /f/ /g/ & /v06/ & /b/ /t\textipa{S}/ /d/ /\textipa{D}/ /f/ /g/ \\ 
 & /hh/ /d\textipa{Z}/ /k/ /l/ /m/ /n/ & & /hh/ /d\textipa{Z}/ /k/ /l/ /m/ /n/ \\
 & /\textipa{N}/ /p/ /r/ /s/ /\textipa{S}/ /t/ & & /\textipa{N}/ /p/ /r/ /s/ /\textipa{S}/ /t/ \\
& /\textipa{T}/ /v/ /w/ /y/ /z/ & & /\textipa{T}/ /v/ /w/ /y/ /z/ \\ 
 /v07/ & /\textipa{Z}/ & /v07/ & /\textipa{Z}/ \\ 
 /sil/ & /sil/ & /sil/ & /sil/ \\ 
 /sp/ & /sp/ & /sp/ & /sp/ \\ 
 /gar/ & /c/ & /gar/ & /c/ \\ 
 \hline 
 	 \end{tabular} 
 \label{tab:lilirsimaps9-10} 
 \end{table} 
 
\begin{table}[!h] 
 \centering 
 	\caption{RMAV speakers 11 and 12} 
 	\begin{tabular} {| l | l || l | l |} 
 	\hline 
	\multicolumn{2}{| c ||}{Speaker 11 $M_{!11}$} & \multicolumn{2}{ c |}{Speaker 12 $M_{!12}$} \\ 
 	Viseme & Phonemes			& Viseme & Phonemes \\ 
 	\hline \hline 
/v01/ & /\textscripta/ /ae/ /\textturnv/ /\textopeno/ /ax/ /ay/ & /v01/ & /\textscripta/ /ae/ /\textturnv/ /\textopeno/ /ax/ /ay/ \\ 
 & /\textipa{E}/ /eh/ /\textrevepsilon/ /ey/ /\textsci\textschwa/ /\textsci/ & & /\textipa{E}/ /eh/ /\textrevepsilon/ /ey/ /\textsci\textschwa/ /\textsci/ \\
 & /iy/ /\textturnscripta/ /\textschwa\textupsilon/ & & /iy/ /\textturnscripta/ /\textschwa\textupsilon/ /\textopeno\textschwa/ \\ 
 /v02/ & /ua/ /\textupsilon/ /\textopeno\textschwa/ & /v02/ & /ua/ /\textupsilon/ \\ 
 /v03/ & /\textscripta\textupsilon/ & /v03/ & /\textschwa/ \\ 
 /v04/ & /\textopeno\textsci/ & /v04/ & /\textscripta\textupsilon/ /\textopeno\textsci/ \\ 
 /v05/ & /\textschwa/ & /v05/ & /b/ /t\textipa{S}/ /d/ /\textipa{D}/ /f/ /g/ \\ 
 & & & /hh/ /d\textipa{Z}/ /k/ /l/ /m/ /n/ \\
 & & & /\textipa{N}/ /p/ /r/ /s/ /\textipa{S}/ /t/ \\
 & & & /\textipa{T}/ /v/ /w/ /y/ /z/ \\ 
 /v06/ & /b/ /t\textipa{S}/ /d/ /\textipa{D}/ /f/ /g/ & /v06/ & /\textipa{Z}/ \\ 
 & /hh/ /d\textipa{Z}/ /k/ /l/ /m/ /n/ & & \\
 & /\textipa{N}/ /p/ /r/ /s/ /\textipa{S}/ /t/ & & \\
 & /\textipa{T}/ /v/ /w/ /y/ /z/ & & \\ 
/v07/ & /\textipa{Z}/ & /sil/ & /sil/ \\ 
 /sil/ & /sil/ & /sp/ & /sp/ \\ 
 /sp/ & /sp/ & /gar/ & /c/ \\ 
 /gar/ & /c/ & & \\ 
 \hline 
 	 \end{tabular} 
 \label{tab:lilirsimaps11-12} 
 \end{table} 

 \begin{table}[!pht] 
\centering 
\caption{AVL2 speakers 1 to 4} 
\begin{tabular}{| l | l || l | l || l | l   } 
\hline 
\multicolumn{2}{| c ||}{Speaker 1 $M_{234}$} & \multicolumn{2}{ c ||}{Speaker 2 $M_{134}$ } \\ 
Viseme & Phonemes 		& Viseme & Phonemes 		\\ 
\hline \hline 
/v01/ 	& 	/\textturnv/ /\textschwa/ /ay/		& /v01/	& 	/\textturnv/ /ay/ /ey/ \\ 
		&	 /ey/ /iy/			&		&	/iy/			\\ 
/v02/ 	& 	/\textschwa\textupsilon/ /uw/ 			& /v02/ 	& 	/\textscripta/ /\textschwa\textupsilon/ /uw/	\\ 
/v03/ 	&	/eh/ 						& /v03/	& 	/\textschwa/ /eh/		\\ 
/v04/ 	& 	/\textscripta/ 				& /v04/ 	& 	/d/ /s/ /t/		\\ 
/v05/ 	& 	/d/ /s/ /t/ /v/ 				& /v05/	& 	/t\textipa{S}/ /l/	\\ 
/v06/ 	& 	/l/ /m/ /n/ 					& /v06/ 	& 	/b/ /d\textipa{Z}/		\\ 
/v07/ 	& 	/d\textipa{Z}/ /p/ /y/ 			& /v07/	& 	/v/ /y/			\\ 
/v08/ 	& 	/k/ /w/ 					& /v08/	& 	/k/ /w/ 		\\ 
/v09/ 	& 	/f/ 						& /v09/ 	& 	/p/			\\ 
/v10/ 	& 	/t\textipa{S}/ 						& /v10/	& 	/z/ 			\\ 
/v11/ 	& 	/b/ 						& /v11/	& 	/m/			\\ 
		&									&		&				\\ 
/sil/ 		& 	/sil/ 						& /sil/	& 	/sil/		 \\ 
/garb/ 	& 	/\textipa{E}/ /\textturnscripta/ /\textopeno/ /r/ /z/ 	& /garb/ 	& /\textipa{E}/ /\textturnscripta/ /\textopeno/ /r/ /f/ /n/	\\ 
\hline \hline 
\multicolumn{2}{| c ||}{Speaker 3 $M_{124}$ } & \multicolumn{2}{ c ||}{Speaker 4 $M_{123}$ } \\ 
Viseme & Phonemes			& Viseme & Phonemes \\ 
\hline \hline 
/v01/ 	&	/\textturnv/ /ay/ /ey/ 		& /v01/ 	& /\textturnv/ /ay/ /ey/ \\ 
&	/iy/ /\textschwa\textupsilon/ /uw/ 	&		& /iy/ /\textschwa\textupsilon/ /uw/ \\ 
/v02/	& 	/\textscripta/				& /v02/	& /\textscripta/ \\ 
/v03/ 	& 	/\textschwa/ /eh/		& /v03/	& /\textschwa/ /eh/ \\ 
/v04/	& 	/d/ /s/ /t/ /v/				& /v04/	& /d\textipa{Z}/ /s/ /t/ /v/\\ 
/v05/ 	& 	/l/ /m/ /n/ 			& /v05/	& /f/ /l/ /n/ \\ 
/v06/ 	& 	/b/ /w/ /y/ 			& /v06/ 	& /b/ /d/ /p/ \\ 
/v07/	& 	/d\textipa{Z}/			& /v07/	& /w/ /y/ \\ 
/v08/	& 	/z/					& /v08/	& /z/ \\ 
/v09/ 	&	/p/ 				& /v09/ 	& /m/ \\ 
/v10/ 	& 	/k/ 				& /v10/ 	& /k/ \\ 
/v11/	& 	/f/ 					& /v11/ 	& /t\textipa{S}/ \\ 
/v12/	& 	/t\textipa{S}/			& 		& \\ 
/sil/ 		& 	/sil/				& /sil/ 	& /sil/ \\ 
/garb/	& 	/\textipa{E}/ /\textturnscripta/ /\textopeno/ /r/ /iy/	& /garb/ 	& ea/ /\textturnscripta/ /\textopeno/ /r/ \\ 
\hline 
\end{tabular} 
\label{tab:l1o_v} 
\end{table}

\end{document}